\documentclass[a4paper,fleqn,usenatbib,longnamesfirst]{mnras}
\usepackage[british]{babel}
\usepackage[T1]{fontenc}
\usepackage{ae,aecompl}
\usepackage{graphicx}
\usepackage{amsmath}
\usepackage{amssymb}
\usepackage{booktabs}		
\usepackage{caption}		
\usepackage{subfig}		
\usepackage[separate-uncertainty=true]{siunitx}	
\usepackage{bm}	
\usepackage{lmodern}	

\providecommand{\abs}[1]{\lvert#1\rvert}
\providecommand{\mean}[1]{\langle #1 \rangle}
\newcommand{\tup}[1]{\textup{#1}}
\newcommand{\msun}{$ \text{ M}_\odot$}
\newcommand{\Mvir}{M_\tup{vir}}
\newcommand{\Rvir}{R_\tup{vir}}
\newcommand{\x}{\times}
\newcommand{\minus}{^{-1}}
\newcommand{\minuss}{^{-2}}
\DeclareSIUnit\kpc{kpc}

\title{On the coherent rotation of diffuse matter in numerical simulations of galaxy clusters}

\author[Baldi et al.]{%
	Anna Silvia Baldi,$^{1}$\thanks{E-mail: anna.silvia.baldi@roma1.infn.it}
	Marco De Petris,$^{1}$
	Federico Sembolini,$^{1,2,3}$\newline
	\newauthor Gustavo Yepes,$^{2,3}$ Luca Lamagna$^{1}$ and Elena Rasia$^{4,5}$
	\\
	\\
	$^{1}$Dipartimento di Fisica, Sapienza Universit\`{a} di Roma, Piazzale Aldo Moro, 5-00185 Roma, Italy\\
	$^{2}$Departamento de F\'{\i}sica Te\'{o}rica, M\'{o}dulo 8, Facultad de Ciencias, Universidad Aut\'{o}noma de
	Madrid, E-28049 Cantoblanco, Madrid, Spain\\
	$^{3}$Astro-UAM, UAM, Unidad Asociada CSIC\\
	$^{4}$INAF, Osservatorio Astronomico di Trieste, via Tiepolo 11, I-34131 Trieste, Italy\\
	$^{5}$Department of Physics, University of Michigan, 450 Church St., Ann Arbor, MI 48109, USA
}

\date{Accepted XXX. Received YYY; in original form ZZZ}

\pubyear{2016}

\begin{document}
\label{firstpage}
\pagerange{\pageref{firstpage}--\pageref{lastpage}}
\maketitle

\begin{abstract}
        We present a study on the coherent rotation of the intracluster medium and dark matter
        components of simulated galaxy clusters extracted from a volume-limited sample of the MUSIC project. 
        The set is re-simulated with three different recipes for the gas physics: $(i)$ non-radiative, $(ii)$ radiative
        without AGN feedback, and $(iii)$ radiative with AGN feedback.
        Our analysis is based on the 146 most massive clusters identified as relaxed, 57 per cent of the total
        sample. We classify these objects as rotating and non-rotating according to the gas spin parameter,
        a quantity that can be related to cluster observations.
        We find that 4 per cent of the relaxed sample is rotating according to our criterion.
        By looking at the radial profiles of their specific angular momentum vector, we find that
        the solid body model is not a suitable description of rotational motions.
        The radial profiles of the velocity of the dark matter show a prevalence of the random velocity dispersion.
        Instead, the intracluster medium profiles are characterized by a comparable contribution from the
        tangential velocity and the dispersion.
        In general, the dark matter component dominates the dynamics of the clusters, as suggested by the correlation
        between its angular momentum and the gas one, and by the lack of relevant differences among the three
        sets of simulations.
\end{abstract}


\section{Introduction}
The dynamics, and the rotational motions in particular, of all matter components
(galaxies, diffuse gas, and dark matter) of galaxy clusters, have not been fully explored yet.
This is mostly due to the complexity of these large systems and to the relatively low statistics available on
their dynamical properties.

From a theoretical point of view there are some mechanisms related to structure formation
that can produce coherent rotational motions of the intracluster medium (ICM) and galaxies,
as well as of dark matter (DM).
A first approach on the study of the origin of global angular momentum in cosmic structures was attempted by
\citet{peebles:angularmomentum}, who proposed the tidal torque from surrounding
matter as a possible cause of a rotation.
A consequence of this interaction is a lognormal distribution of the
spin parameters of DM haloes \citep{catelan:angmomevolution}.
An alternative mechanism that can give origin to a large-scale rotational motion is the occurrence of
off-axis merging events, \citep{ricker:offaxismergers98, ricker:offaxismergers} that can be induced by tidal
torque itself \citep{roettiger:offaxisrotation}.

The presence of rotational motions in clusters can affect some of their structural and
evolutionary properties.
The mass function derived from galaxy cluster studies is one of the most effective
approaches to obtain cosmological parameters \citep[see e.g.][]{vikhlinin:clusterparam,planck:clustercount}.
Its determination, however, is limited by the systematics affecting the measure of the cluster total mass.
In this context, the understanding of gas motions is of fundamental importance to solve the
problem of the mass bias due to the lack of hydrostatic equilibrium \citep[see e.g.][]{biffi:hse}.
Non-thermal contributions to the gas pressure, indeed, might come from turbulence, but also
from coherent rotation.
In \citet{nipoti:rotmhd}, the effect of ICM rotation on the
magneto-rotational stability in the cool core of relaxed clusters is investigated.
Another process that may be influenced by a rotation of the gas is the accretion of matter onto the central
cD galaxies in clusters \citep{tutukov:rotevolution}.

From an observational point of view, multi-wavelength techniques applied to different observables
can be used for the study of the dynamics of clusters through the exploration of ICM and galaxy members.
Observations are still at a primordial stage, nowadays; nevertheless some
attempts to measure the rotational velocity of ICM and galaxies have been made.
A method that is used to investigate rotational motions,
and to possibly classify clusters as rotating, consists in checking
the presence of velocity gradients in different regions of the clusters with respect to the centre.
\citet{dupke:perseus}, using Doppler shift of X-ray emission lines from \textit{ASCA} observations of the ICM in
Perseus cluster, have found a velocity gradient in the outer regions which is consistent with
a rotational velocity of almost \SI{1000}{\km\per\second} at 90 per cent confidence level.
Recently the central regions ($ r \lesssim \SI{100}{\kpc}$) of the same cluster have been observed
with the \textit{Hitomi} satellite \citep{hitomi:perseus}. Data reveal a gradient of the velocity projected on the line
of sight of about \SI{150 \pm 70}{\km\per\second}, and a velocity dispersion that is compatible with a relatively
low amount of turbulent motions, suggesting a quiescent dynamics.
Another recent application of X-ray spectroscopy to explore gas motions can be found in \citet{liu:ccd,liu:ccd2},
where \textit{Chandra} data have been used for the study of bulk motions in disturbed clusters.
Still referring to observations of the diffuse baryonic component, a coherent rotation of ICM can also
be investigated in the millimetric band by observing the induced kinematic Sunyaev--Zel'dovich (SZ)
contribution to cosmic microwave background temperature fluctuations. This challenging
target has been explored by \citet{cooray:ksz} assuming a solid body model for a cluster,
and described analytically by \citet{chluba:ksz}.
A first study on the detectability of this signal is reported in \citet{sunyaev:turbulence}, and it will be the
topic of a forthcoming paper in preparation.
The aforementioned analysis of velocity gradients
has been also applied to spectroscopic observations of galaxies
in the optical band \citep[see e.g.][]{denhartog:abell,biviano:coma,hwang:sdssrotating}.
A slightly different approach is adopted in \citet{tovmassian:galaxies}, as the rotational state is inferred
from the spatial distribution of member galaxies having higher or lower velocities with respect to the
mean global velocity. With this criterion 26 per cent of the analysed clusters is found to be rotating.
\citet{plionis:rotation} have recently proposed a variant of the velocity gradients method,
identifying rotating clusters from the projection of the velocity of single galaxies along the line of sight.
In a total sample of $ \sim 50 $ analysed clusters, they find a fraction of 35 per cent rotating candidates with
galaxy rotational velocities of the order of thousands of \si{\km\per\second}.
A possible comparison of the results from the observational approaches applied to the diffuse gas
and those applied to the discrete galaxies, could be a more robust way to establish the presence of a
rotation in clusters.

Cosmological N-body simulations are useful tools to characterize in more detail the dynamics within
cosmic structures.
Mock X-ray signals from simulated clusters have been produced to investigate the indications of a possible
rotation. For example, \citet{roettiger:offaxisrotation} simulated a cluster with the same characteristics of
Abell 3266, and applied an off-axis merger that produced a rotation associated to bulk flows of about
$ \SI{800}{\km\per\second} $.
The study of the ellipticity of the isophotes in simulated X-ray surface brightness maps has been proposed
as another possible way to infer the dynamics of a cluster, as it could be a sign of a rotation
\citep{fang:rotationandturbulence,bianconi:articolo}, even if there is not a common agreement
\citep[see e.g.][]{biffi:velocity}.
There are different works in literature where the angular momentum properties of large samples of
synthetic objects are investigated
\citep[e.g.][]{barnes:angularmomentum,efstathiou:velocity,bullock:articolo, vandb:angularmomentum, 
sharma:angularmomentum}, mostly with the application to DM haloes.
With the improvement of the models adopted to simulate the evolution of gas and astrophysical
processes, the ICM dynamics has also been analysed
\citep[see e.g.][]{rasia:ICMDMdensity, faltenbacher:motions, fang:rotationandturbulence, lau:motions,
biffi:velocity, lau:weighing}, recently also with the prescription for AGN feedback
\citep[e.g. in][]{biffi:phox,nagai:motions}.
From a statistical analysis of simulated objects, it has emerged that rotational motions seem to predominate
in the innermost regions of disturbed clusters rather than in relaxed ones \citep{biffi:velocity}, while in
outer regions relaxed clusters show a larger contribution from gas rotation instead
\citep{fang:rotationandturbulence,lau:motions}.
\newline \indent
In this work we extract for the first time information on spin, angular momentum and velocity properties
of ICM and DM in a volume-limited sample of \textit{massive clusters of galaxies} from MUSIC simulations,
for which a non-radiative model and two radiative models for gas physics have been used.
To this scope we compute the specific angular momentum and the velocity profiles along the cluster radius,
also looking for possible correlations between DM and ICM.
We adopt in particular a couple of recently proposed models with which we compare the radial profiles
of the tangential velocity, that we derive from the specific angular momentum.
\newline \indent
This paper is organized as follows.
In section~\ref{sec:musicclusters} we describe the main characteristics of the sample of simulated clusters.
In section~\ref{sec:spinparameter} we perform the analysis of the spin parameter of these clusters.
Section~\ref{sec:angmomprofiles} is devoted to the description of the radial profiles of the angular
momentum of DM and ICM in relaxed clusters and shows the corresponding behaviours.
The main results of the analysis of the tangential velocity and its dispersion are presented in
section~\ref{sec:velocities}, together with a new model that we propose to describe the ICM rotational motions.
In section~\ref{sec:componentsbehaviour} we show the comparison
of the angular momentum at virial radius between the DM and gas components.
Finally, we summarize our results in section~\ref{sec:conclusions}.

\section{The dataset}
\label{sec:musicclusters}
\subsection{MUSIC}
The MUSIC\footnote{\texttt{http://music.ft.uam.es}} dataset is one of the largest catalogues of hydrodynamic
simulations of galaxy clusters, with more than 700 clusters and 2000 groups of galaxies \citep{sembolini:music1}.
It consists of two sets (named MUSIC-1 and MUSIC-2) of re-simulated objects extracted from two
large-volume N-body simulations.
In this paper, we focus on the MUSIC-2 sample containing a large statistic of
massive objects.
The systems were selected from the (1$h\minus$ Gpc)$^3$ volume of the MultiDark simulation
\citep{prada:multidark}, performed using the best-fitting cosmological parameters from WMAP7+BAO+SNI
($\Omega_m = 0.27 $, $\Omega_b = 0.0469 $, $\Omega_{\Lambda} = 0.73 $, $\sigma_8 = 0.82 $, $n = 0.95 $,
$h = 0.7$) \citep{komatsu:wmap}.

Once the clusters were identified, the zooming technique by \citet{klypin:zooming} was adopted to create
new initial conditions. These enabled re-simulations at higher mass resolution of the spherical regions
around the cluster centres with radii equal to $ 6 h\minus $ Mpc.
The new sets were carried out with the parallel TreePM+SPH GADGET-3 code which includes the
entropy-conserving implementation of smoothed particle hydrodynamics (SPH) \citep{springel:gadget}.
Three re-simulation sets were produced accounting for different models to describe the baryon physical processes.
We will refer to NR for the non-radiative subset, to CSF for the run including cooling, 
star-formation and stellar feedback, and to AGN to the simulation that, further, adds the AGN feedback.
The DM particle mass is set to $ m_\tup{DM}=\num{9.0e8} h\minus $\msun, while the gas mass particle is
equal to $ m_\tup{gas} = \num{1.9e8} h\minus $\msun\ in the NR set, and it is variable in the radiative simulations.
Even in this case, however, $ m_\tup{gas} $ is still of the order of \num{e8}$h\minus$\msun.

When modelling the sub-grid physics of our CSF and AGN subsets, we accounted for the effects of radiative cooling,
UV photoionization, star formation and supernova feedback, including the effects of strong winds from supernovae,
as described in the \citet{springel:starformation} model.
Stars are assumed to form from cold gas clouds on a characteristic timescale $t_{\star}$, and a stellar mass fraction 
$\beta$ = 0.1 (consistent with assuming an Universal Salpeter IMF with a slope of $-1.35$) is instantaneously
released due to supernovae from massive stars ($M > 8$\msun).
In addition to this mass injection of hot gas, thermal energy is also released to the interstellar medium
by the supernovae.
The number of collisionless star particles spawned from one SPH parent gas particle is fixed to 2.
Kinetic feedback is also included: gas mass losses due to galactic winds, $\dot{M}_\tup{w}$, is assumed to be
proportional to the star formation  rate $M_\tup{SFR}$, i.e. $\dot{M}_\tup{w}$ = $ \eta M_\tup{SFR}$
with $\eta$ = 2.
Additionally, the wind contains a fixed fraction $\epsilon$ = 0.5 of the total supernova energy. 
SPH particles near the star formation region are subject to enter in the wind in an stochastic way,
given an isotropic velocity kick of $v$ = \SI{400}{\kilo\meter\per\second}.
The simulations including AGN feedback have been carried with the same version of the GADGET-3 code
that has been used for the simulations presented in \citet{planelles:agn}.
This model is based on the original implementation by \citet{springel:bh} (SMH model),
with feedback energy released as a result of gas accretion onto supermassive black holes (BH).
In this AGN model, BHs are described as sink particles, which grow their
mass by gas accretion and merging with other BHs.
The seeding of BH particles has been modified with respect to the original SMH model,
and occurs only in haloes where star formation took place. A minimum mass of
$5\times10^6 h \minus$\msun\ is assumed for a friends-of-friends (FoF)
group of star particles to be seeded with a BH particle. Seeded BHs are located at the potential minimum
of the FoF group, instead of at the density maximum, as originally implemented by SMH. The pinning of the
BH is regulated at each time-step to avoid advection.
In this way a BH particle remains within the host galaxy, when this becomes a satellite of a larger halo.
A more strict momentum conservation during gas accretion and BH merger is also set.
Two BHs now merge when they are located at a distance from each other that is less than the
gravitational softening and when their relative velocity is less than half of the sound speed.
Finally, the AGN feedback is provided via thermal energy to the surrounding gas particles.
Eddington-limited Bondi accretion produces a radiated energy which corresponds to a fraction $ \epsilon_r = 0.1 $
of the rest-mass energy of the accreted gas. A fraction of this radiated energy is thermally coupled to the
surrounding gas with feedback efficiency $ \epsilon_f = 0.1 $.
This parameter is regulated to reproduce the observed relation between the BH mass and stellar mass of the
hosting halo \citep{ragone:agn}.
Special care is devoted to the treatment of  multi-phase and star forming particles to avoid loosing the
AGN energy \citep[see][for details]{planelles:agn}.
No mechanical feedback is implemented in these runs, therefore jets and raising bubbles are not described.
The transition from a `quasar' phase to a `radio' mode of the BH feedback
happens when the accretion rate onto the BH becomes smaller than 1 per cent of the Eddington accretion
\citep[see also][]{sijacki:agn,fabjan:agn}.
At that instant, the efficiency of the AGN feedback is enlarged by a factor of 4.

\subsection{Selected sample}
We analyse 258 simulated massive clusters with virial masses $ \Mvir > \num{5e14} h\minus$\msun\ at $z = 0$,
extracted from the MUSIC-2 subset.
A first reason for the choice of this mass range is that MUSIC-2 is a complete sample in mass:
all the massive objects above a given mass threshold (which varies with redshift) formed in the MultiDark
parent simulation have been re-simulated.
A second reason lies in the fact that we expect that a possible rotation, which here is investigated
by means of the properties of the angular momentum, would be more likely observed/measured in the most
massive clusters.
Each cluster is analysed in the three aforementioned different flavours (NR, CSF and  AGN).
Clusters in the CSF and NR datasets have been already employed to study SZ scaling
relations~\citep{sembolini:music1, sembolini:music3} and X-ray properties~\citep{biffi:music2}.
The reliability of our code was tested in comparison with different gas-dynamical codes to
study the consistency between simulated clusters
modelled with different numerical and radiative models \citep[see][]{nifty:1, nifty:3, nifty:4, nifty:2}.

For our analysis we have made a further classification on the basis of the relaxation and the rotation state,
as described in detail in section~\ref{sec:angmomprofiles}.
All the useful informations such as particle masses, haloes centre of mass and three-dimensional 
velocities were retrieved using the Amiga Halo Finder~\citep[AHF,][]{knebe:ahf}.

\section{Spin parameter analysis}
\label{sec:spinparameter}
The rotational state of a halo can be quantified by calculating the spin parameter $ \lambda $.
We adopt the simplified expression from \citet{bullock:articolo} derived from the total angular
momentum $ L_\tup{tot} $:
\begin{equation}
	\label{eqn:lambdabullock}
	\lambda_\tup{tot} = \frac{L_\tup{tot}}{\sqrt{2} v_\tup{circ} M_\tup{vir} R_\tup{vir}}
\end{equation}
where $ v_\tup{circ} $ indicates the circular velocity as calculated at virial radius, $ v_\tup{circ}=\sqrt{G \Mvir/\Rvir} $.
We remind that other definitions are present in literature \citep[see][]{peebles:angularmomentum,
bullock:articolo, yepes:marenostrum, bryan:spin}. However, we prefer to use equation\eqref{eqn:lambdabullock}
for its simplicity, and because we can use it to express the spin parameter of each single matter component
(ICM or DM), that we will identify as $ \kappa $ in the following:
\begin{equation}
	\label{eqn:lambdagasDM}
	\lambda_{\kappa} = \frac{L_{\kappa}}{ \sqrt{2 G M_\tup{vir} R_\tup{vir}} \ M_{\kappa}}
\end{equation}
 \citep{yepes:marenostrum}.
We find that the values of the DM spin parameter are very close to those derived from the total angular momentum.
This is not surprising since the mass of the DM component dominates over the baryonic one.
The relation between $ \lambda_\tup{tot} $ and $ \lambda_\tup{DM} $ is found to be linear, with a slope
close to unity for all the three analysed subsets (NR, CSF and AGN).
\begin{figure}
	\centering
	\includegraphics[width=0.50\textwidth]{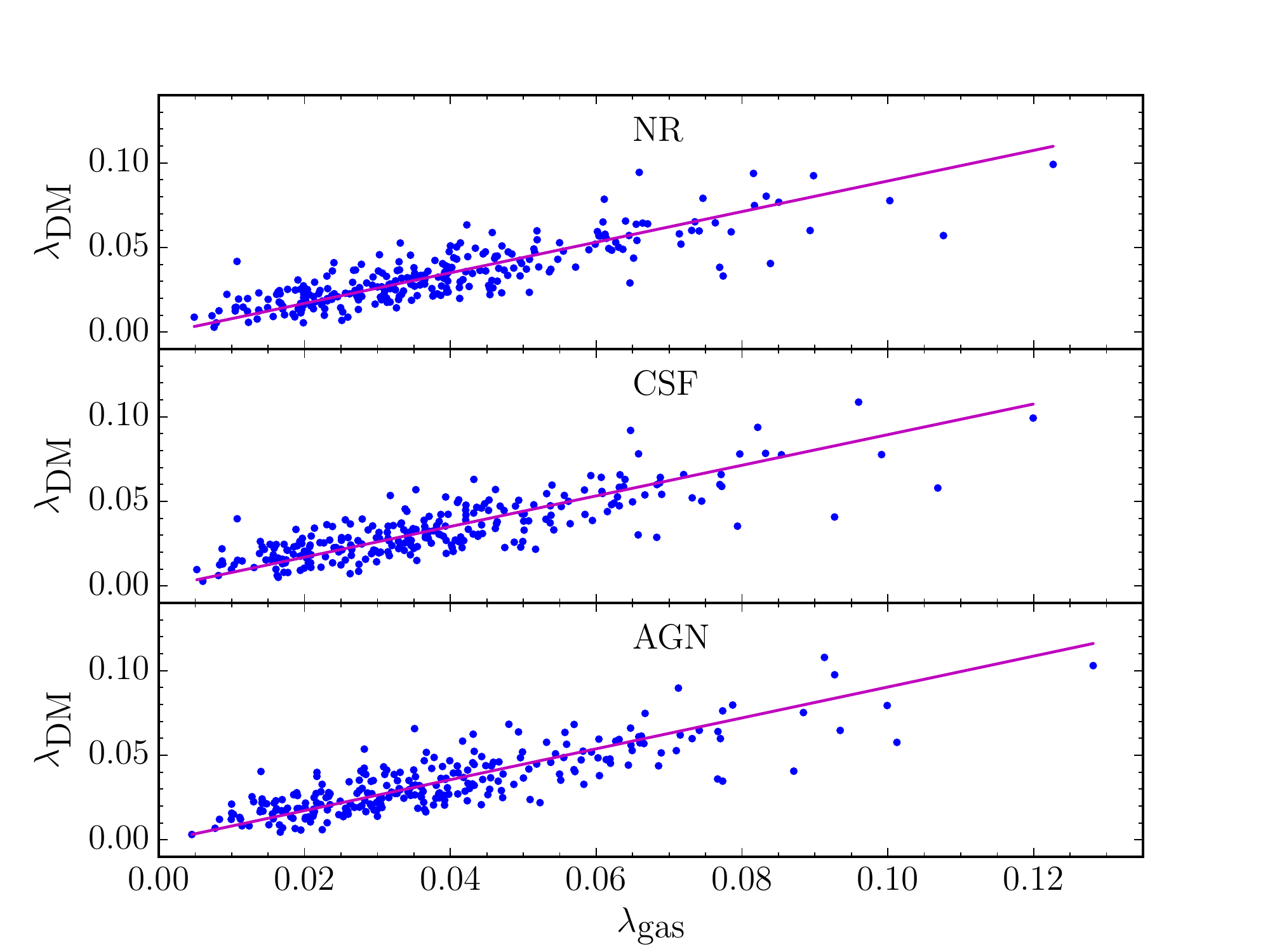}
	\caption{\small Relation between the spin parameters of the DM and ICM
		components for all the clusters in the sample. Solid magenta lines represent the robust linear fits.}
	\label{fig:lambdas}
\end{figure}
\begin{table}
	\centering
	\caption{\small  Values of the correlation coefficient ($ c_\tup{corr} $) and of the parameters for
		the linear fits shown in Fig.~\ref{fig:lambdas}, performed with the bisector method;
		$ a $ indicates the slope, while $ b $ indicates the zero-point.}
	\begin{tabular}{cccc}
		\toprule
		Dataset & $ c_\tup{corr} $ & $ a \pm \sigma_a $ & $ b \pm \sigma_b$\\
		\midrule
		NR & $ 0.84 $ & $ 0.90 \pm 0.04 $ & $ -0.001 \pm 0.002 $\\
		CSF & $ 0.83 $ & $ 0.91 \pm 0.05 $ & $ -0.001 \pm 0.001 $\\
		AGN & $ 0.83 $ & $ 0.91 \pm 0.04 $ & $ -0.001 \pm 0.001 $\\
		\bottomrule
	\end{tabular}
	\label{tab:lambdacompfit}
\end{table}

\begin{figure}
	\centering
	\includegraphics[width=0.50\textwidth]{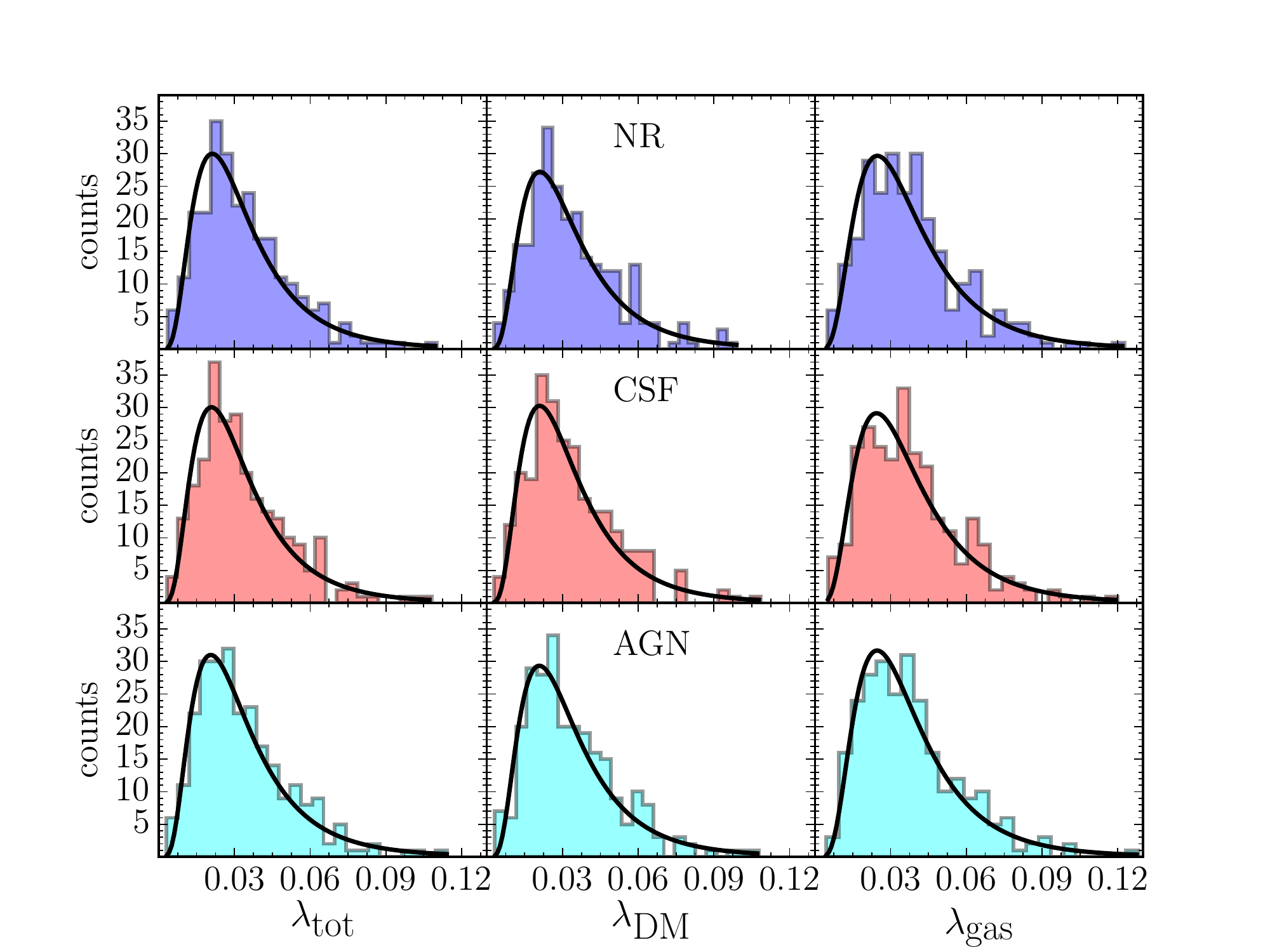}
	\caption{\small Distributions of the spin parameters of each matter component (along columns)
		for the three analysed subsets (along rows). Solid lines represent the fits to the lognormal distribution.}
	\label{fig:lambdadistributions}
\end{figure}

In order to explore a possible correlation between the angular momentum of DM and gas,
we compare the corresponding spin parameters. A clear linear relation between the two, with a slope
$ a \sim 0.90 $ is shown in Fig.~\ref{fig:lambdas}, and the full set of parameters obtained
from a robust fit to the data using the bisector method \citep{isobe:bisector} is listed in 
Table~\ref{tab:lambdacompfit}.\newline
Fig.~\ref{fig:lambdadistributions} shows the distributions of the spin parameters for gas, DM and total
matter components of the clusters in our sample, for the three different flavours.
We can see from these results that a lognormal distribution
\begin{equation}
	\label{eqn:lognormaldistribution}
	P(\lambda) d \lambda = \frac{1}{\lambda \sqrt{2 \pi }\sigma}
	\exp \left(- \frac{\ln^2(\lambda/\lambda_0)}{2 \sigma^2} \right) d \lambda
\end{equation}
is a valid description of the spin parameters of all the matter components.
The scale parameter $ \lambda_0 $ and the shape parameter $ \sigma $ derived from our sample of clusters
are reported in Table~\ref{tab:lambdaparameters}.
As expected, the similarity between total and DM values is also evident from these results.
When comparing our values of the parameters with other works,
\citep{barnes:angularmomentum, bullock:articolo, vandb:angularmomentum, %
sharma:angularmomentum, yepes:marenostrum,maccio:spin,bryan:spin}
we find a general agreement. It is worth to stress that these works
refer to galactic or proto-galactic haloes, with the exception of \citet{maccio:spin}, where
objects on different scales are considered (from galaxies to clusters), and of \citet{yepes:marenostrum},
which include clusters of galaxies of the MareNostrum gas-dynamical simulation.
Therefore, this result suggests that the shape of the distribution does not vary significantly from galaxies
to clusters.
The values of $ \lambda_\tup{0,gas} $ are typically larger than those of the DM (by 13 per cent in our case),
suggesting more rotational support. This may be because tidal interactions with surrounding large-scale
structures have had more time to apply torques to the gas accreted at a later time.
\begin{table}
	\centering
	\caption{Values of the parameters $ \lambda_0 $ and $ \sigma $ from the lognormal fits to the spin
		parameter distributions for DM, gas and the total matter components of the analysed sample.}
	\begin{tabular}{llllll}
		\toprule
		$\lambda_\tup{0,DM}$ & $ \sigma_\tup{DM} $ & $ \lambda_\tup{0,gas}$ & $ \sigma_\tup{gas} $ &
		$ \lambda_\tup{0,tot}$ & $ \sigma_\tup{tot} $\\
		\midrule
		\multicolumn{6}{c}{NR}\\
		0.0289 & 0.5674 & 0.0333 & 0.5470 & 0.0292 & 0.5655\\
		\midrule
		\multicolumn{6}{c}{CSF}\\
		0.0288 & 0.5638 & 0.0330 & 0.5489 & 0.0287 & 0.5641\\
		\midrule
		\multicolumn{6}{c}{AGN}\\
		0.0289 & 0.5755 & 0.0330 & 0.5416 & 0.0289 & 0.5810\\
		\bottomrule
	\end{tabular}
	\label{tab:lambdaparameters}
\end{table}
\section{Radial profiles of the angular momentum}
\label{sec:angmomprofiles}
After characterizing the spin computed at the virial radius, we move here to study the variations of
the specific angular momentum vector along the cluster radius. We produce
radial profiles describing of its modulus and orientation with respect to the angular momentum
computed at the virial radius.
We consider 15 concentric spheres with radius increasing logarithmically,
from $ r = 0.05 R_\tup{vir} $ to $ r = \Rvir $.
We do not impose any condition on the number of particles in each sphere,
but we would like to remark that the minimum amount of particles is always above \num{e4}, i.e.
enough to lead to robust results.
For the $ i $-th sphere, the modulus of the specific angular momentum $j(< r_i) = \abs{\bm j(< r_i)}$
is estimated as
\begin{equation}
\label{eqn:lspec}
j(< r_i) = \frac{\abs{\bm L(< r_i)}}{M(< r_i)} =
\frac{\abs{\sum\limits_{k}^{N_i} \bm r_k \x m_k \bm v_k}}{\sum\limits_{k}^{N_i} m_k}
\end{equation}
where $ N_i $ is the number of particles inside the $ i $-th sphere, $ \bm r_k $ is the position of the $ k $-th particle 
relative to the centre of mass, $ m_k $ is its mass, and $ \bm v_k $ is its velocity vector
subtracted by the velocity of the centre of mass.\\
The angle $ \alpha(< r_i) $ indicates the variation of the direction of the angular
momentum with respect to its orientation at virial radius, and is defined as
\begin{equation}
\label{eqn:angolomomenti}
\alpha (< r_i)= \arccos \left[\hat j (< r_i) \cdot \hat j(< \Rvir) \right]
\end{equation}
being $ \hat j(< r) = \bm j(< r)/j(< r) $.\\
In the case of a simple solid body rotation, $ \alpha (< r_i) $ is expected to be always null along the cluster radius.

We aim at analysing the coherent rotation of the ICM only in morphologically relaxed clusters, to
avoid the influence of outliers in the velocity and in the angular momentum distributions,
due to the presence of mergers or of any large substructure within the virial radius.
To evaluate the cluster \textit{dynamical state}, we use two of the most adopted indicators
present in literature
\citep[see e.g.][]{neto:relaxation,power:relax,killedar:baryonslensing,meneghetti:clash,sembolini:music3,
	klypin:relaxation,biffi:hse}.
The first indicator is the spatial offset between the density peak position and the centre of mass position,
normalized to the virial radius, $\Delta r = \abs{r_{\delta} - r_\tup{CM}}/\Rvir$.
The second indicator is the ratio between the mass of the largest
substructure within the virial radius, and the cluster virial mass, $ M_\tup{sub}/\Mvir $
\citep[see for instance][for a more detailed discussion]{sembolini:music3}.
The assumed threshold values are $ \Delta r = 0.10  $ for the centre of mass offset,
according to~\cite{donghia:deltar2}, and $ M_\tup{sub}/\Mvir = 0.10 $ for the largest substructure mass to virial
mass ratio~\citep{ascasibar:vtan,sembolini:music3,meneghetti:clash}.
If the values of these indicators are below the respective thresholds, the clusters are classified
as relaxed, otherwise they are disturbed and they are not considered in this analysis.
With these criteria we select 146 relaxed clusters, corresponding to the 57 per
cent of the total sample.
This fraction is consistent with results from both observational data \citep[see e.g.][]{rossetti:planckrelax}
and analyses on the morphology of MultiDark simulated clusters \citep{vega:shape}.

To define the \textit{rotational state} of a relaxed cluster we consider the value of the spin parameter 
of the gas as the discriminant indicator, since it quantifies the contribution of the gas rotational energy
to the total energy of the cluster, by definition.
We classify a cluster as rotating if it satisfies the condition $ \lambda_\tup{gas} > \lambda_\tup{gas,crit} $,
where $ \lambda_\tup{gas,crit} $ is the threshold that separates the total sample in two sub-samples
showing distinguishable profiles of the tangential velocity (see appendix~\ref{sec:appendix} for details).
In our case $ \lambda_\tup{gas,crit} = 0.07 $, according to which about 4 per cent of the relaxed cluster
sample can be classified as rotating.
In separating the population of the relaxed and rotating clusters, the corresponding conditions have been
imposed to be valid for all the three subsets (NR, CSF and AGN).
In order to verify whether the most massive clusters have the largest rotational support,
the correlation between $ \Mvir $ and $ \lambda_\tup{gas} $ has been investigated. Interestingly, we find
that the clusters classified as rotating are not the most massive objects in the sample.
It is worth to stress, however, that this sample contains all the clusters more massive than
\num{5e14} $h\minus$\msun\ that have formed within (1$ h\minus $ Gpc)$^3$ volume, with the adopted
cosmological model as described in section~\ref{sec:musicclusters}.
This leads to an intrinsically limited statistics of objects having large masses, that may contribute in finding
a relatively small number of massive rotating clusters.

\begin{figure}
	\centering
	\subfloat[direction] %
		{\includegraphics[width=0.50\textwidth]{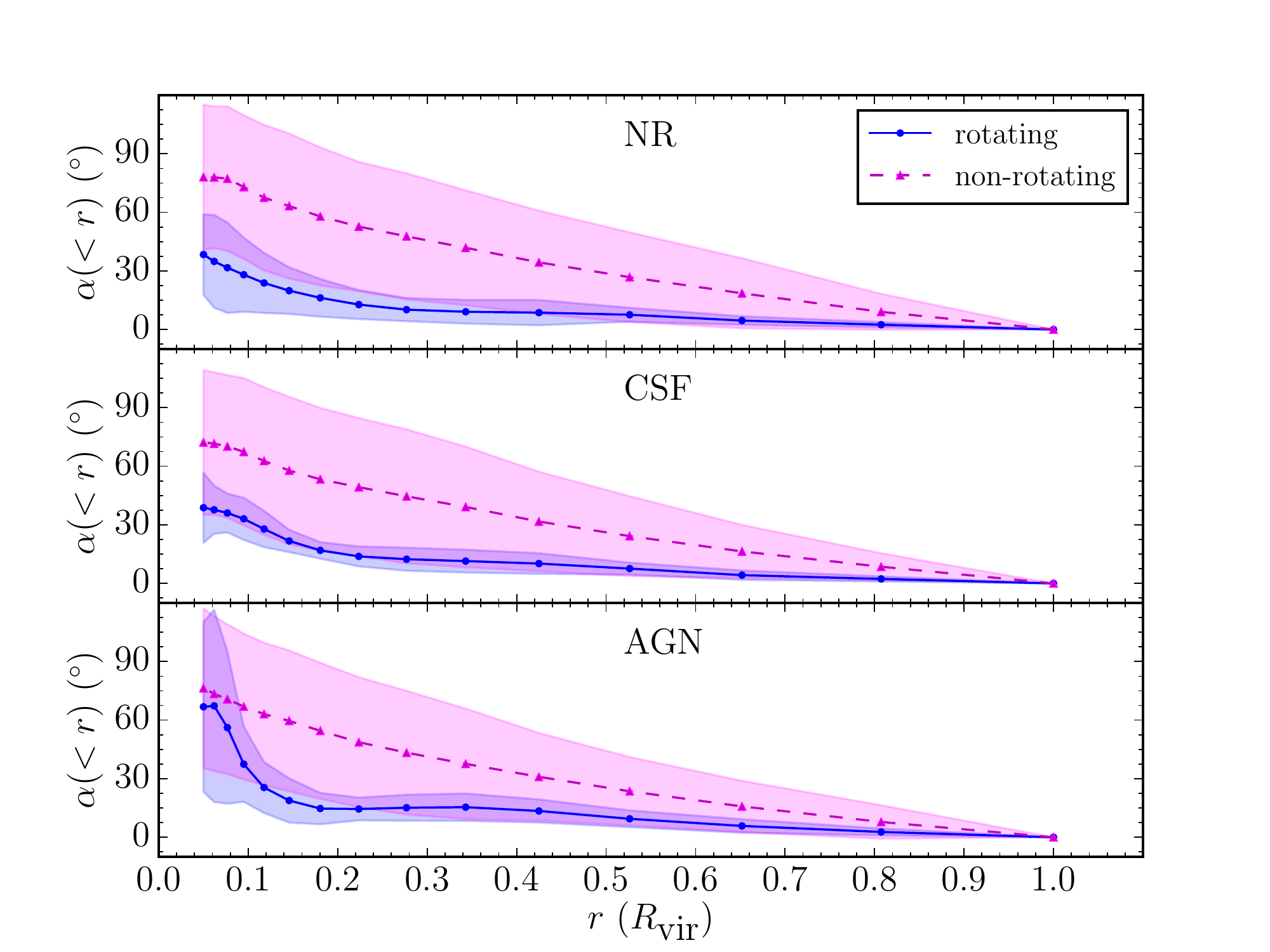}\label{fig:2rotgas_alpha}}
		\quad
	\subfloat[modulus] %
		{\includegraphics[width=0.50\textwidth]{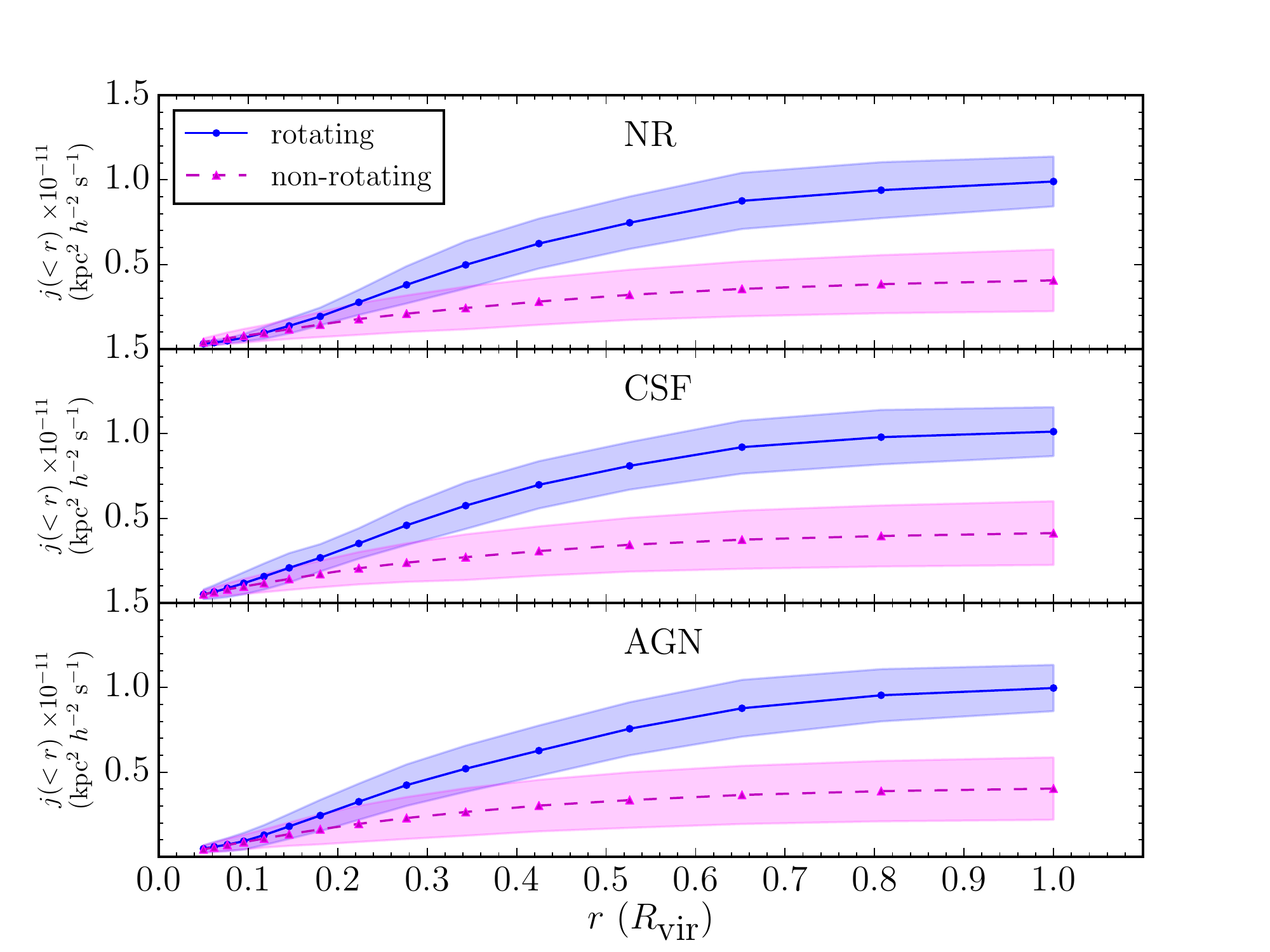}\label{fig:2rotgas_j}}
	\caption{\small Radial profiles of the orientation (upper panel) and the modulus (lower panel) of the gas specific
		angular momentum vector for the two populations of relaxed rotating and non-rotating clusters.
		The points in the plots represent the mean values for each population, and the shaded bands
		indicate the 1$ \sigma $ scatter with respect to the mean.}
\end{figure}

The ICM mean profiles of $j(< r)$ and $\alpha(< r)$ for the two classes of rotating and non-rotating
clusters, are shown in Fig.~\ref{fig:2rotgas_alpha} and Fig.~\ref{fig:2rotgas_j} respectively;
the profiles for DM are characterized by similar features.
Some general trends are evident.
On average, the direction of $ j(< r) $ reaches more than \SI{60}{\degree} in the core of non-rotating clusters
and it is still above \SI{30}{\degree} at half of the virial radius. It is also noticeable that several objects register
a variation equal and greater than \SI{90}{\degree} from the core to the outskirts.
The rotating clusters show a much smaller variation: for $ r \gtrsim 0.3 \Rvir $ we find that $ \alpha(< r) $ is
less than $ \sim \SI{20}{\degree} $, pointing out that the orientation is almost fixed.
The profiles of the modulus increase from the centre up to the virial radius and flatten in the
outskirts, reaching larger values in the rotating clusters, as expected.
A similar behaviour has been also found in~\citet{bullock:articolo}, who analysed only
the DM component of galaxy-sized haloes.
They found a power-law relation of the type $ j(< r) \ \propto \ r^\beta $ with $\beta = 1.1 \pm 0.3 $.
In our sample we perform a similar fit to a power-law $ j(< r)/j(< R_\tup{vir}) \ \propto \ r^{\beta} $
to the profiles over spherical shells, normalized at virial radius for the gas and DM
components in the rotating clusters. The mean values of $ \beta $ with the corresponding standard deviations
are listed in Table~\ref{tab:jnormparameters}.
\begin{table}
	\centering
	\caption{\small Mean values and standard deviations of the power-law exponent $ \beta $ as
		derived from the fits of $ j(< r)/j(< R_\tup{vir}) $ for the rotating clusters.}
	\begin{tabular}{ccc}
		\toprule
		Dataset & \multicolumn{2}{c}{$ \beta \pm \sigma_{\beta} $}\\
		\midrule
		& gas & DM \\
		NR &  $ 0.78 \pm 0.13 $ & $ 0.96 \pm 0.13 $\\
		CSF &  $ 0.69 \pm 0.10 $ & $ 0.92 \pm 0.15 $\\
		AGN &  $ 0.75 \pm 0.13 $ &  $ 0.90 \pm 0.14 $\\
		\bottomrule
	\end{tabular}
	\label{tab:jnormparameters}
\end{table}
The power-law profiles of the DM are in agreement with \citet{bullock:articolo}, while
the values of the gas are 20 per cent lower.

Our results lead to the conclusion that the coherent rotational motions of ICM and DM in our cluster sample
are not properly described by a simple solid body model.
A further confirmation of this fact is given by the angles between the angular momentum vector
and the three semi-axes describing the ellipsoids that approximate the shape of the matter distribution
of gas and DM. Considering only the case of rotating clusters, in fact, these angles range from tens up to
\SI{180}{\degree}, suggesting a misalignment that is not compatible with a rigid rotation.

Finally, we compute $ \alpha(< r) $ and $ j(< r) $ for the total angular momentum, and comparing the
profiles with those obtained for the gas and the DM we found a very close similarity with the latter,
reflecting the predominance of this component on the ICM.
The dominating role of DM in the cluster dynamics can also be inferred from the lack of
significant differences between the results obtained for the three physical flavours of the simulations used to
describe the ICM.
The only marked difference is the higher average value of gas $ \alpha(< r) $, associated with a significant
dispersion in the core of the AGN runs ($ r < 0.1\Rvir $). In this case the AGN feedback likely influences the
motion of the gas that, receiving extra energy from the central source, buoyantly raises without any
pre-selected orientation. The effect in real clusters might be even more intense for the presence of the
AGN jets which are not included in our model.

\section{Radial profiles of the velocity}
\label{sec:velocities}
We study the radial profiles of the tangential velocity (or rotational velocity) of gas and DM particles,
expressed as:
\begin{equation}
	\label{eqn:vtang}
		v_\tup{tan}(r_i) = \frac{\abs{\sum\limits_{k}^{N_i} \frac{\bm r_k \times m_k \bm v_k}{\abs{\bm r_k}}}}
	{\sum\limits_{k}^{N_i} m_k} \ ,
\end{equation}
where the sums are extended to the $ N_i $ particles located within the 15 spherical shells enclosed
between the radii $ r_{i-1} $ and $ r_i $, and not to the spheres used above.
In this way we get the local values of the tangential velocity, that we use to test possible rotational behaviours.
We can derive the tangential velocity from the specific angular momentum, by approximating
equation~\eqref{eqn:vtang} using $ v_\tup{tan}(r_i) \sim \mean{\abs{\bm j(r_i)}}/r_i$.
In the second term, the contribution from random turbulence motions is averagely null by definition
\citep{ascasibar:vtan}, thus the average angular momentum computed in a given shell considers only the
contribution from rotational coherent motions.

The velocity component associated to macroscopic random motions will be referred hereafter as turbulence,
denoted with $ v_\tup{turb} $.
We quantify it from the dispersion with respect to the average tangential velocity as in equation~\eqref{eqn:vtang}:
\begin{equation}
	\label{eqn:vturb}
	\resizebox{0.92\columnwidth}{!}{ %
	$v_\tup{turb}(r_i) =
		\left[ %
			\sum\limits_k^{N_i} m_k
			\left(\frac{\abs{\bm r_k \times \bm v_k}}{\abs{\bm r_k}} - v_\tup{tan}(r_i) \right)^2 /
			\sum\limits_k^{N_i} m_k
		\right]^{\frac{1}{2}}$
	} %
	,
\end{equation}
where the sums are extended also here to spherical shells for gas and DM particles.

\begin{figure}
	\centering
	\includegraphics[width=0.50\textwidth] {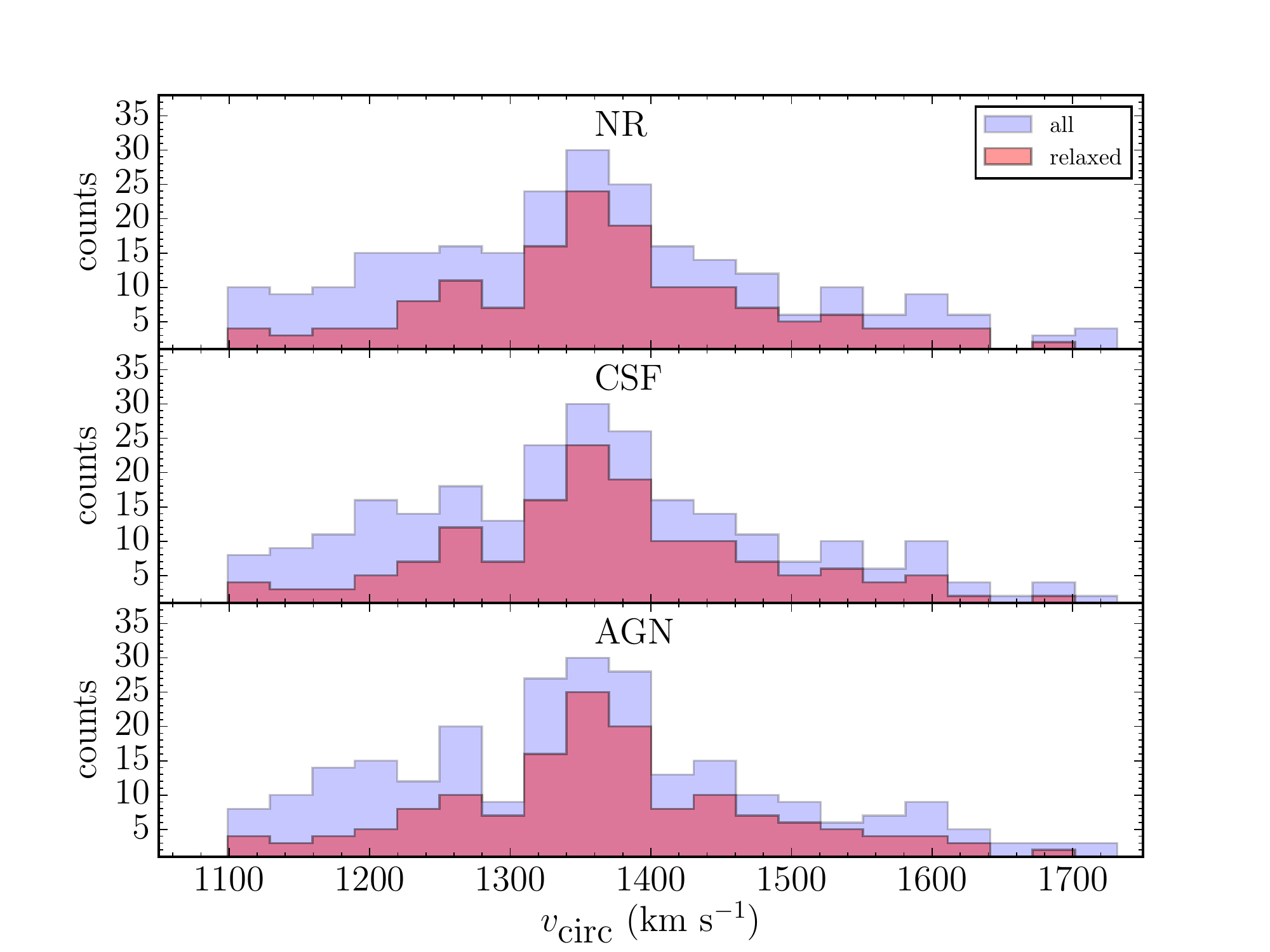}
	\caption{\small Distribution of the circular velocity as calculated at the virial radius for the clusters in the
		analysed sample, with the discrimination of the relaxed ones.}
	\label{fig:histvcirc}
\end{figure}

Both velocity profiles (equations~\eqref{eqn:vtang} and~\eqref{eqn:vturb}) are
normalized to the circular velocity $ v_\tup{circ} $ of the corresponding cluster at $ \Rvir $.
In our sample we have an average value of $ \mean{v_\tup{circ}} = \SI{1365(145)}{\km\per\second} $.
The distribution is shown in Fig.~\ref{fig:histvcirc}.
Due to the tight correlation between the total mass and the circular velocity \citep{evrard:dmvirialscaling},
these distributions emphasize that there is no mass segregation for relaxed clusters.

The mean profiles of the tangential and turbulent velocity have been calculated for
both classes of clusters (rotating and non-rotating) introduced in section~\ref{sec:angmomprofiles},
for all the physical flavours.
Since the tangential velocity is derived from the specific angular momentum, whose direction changes
significantly along the radius (Fig.~\ref{fig:2rotgas_alpha}), we multiply its values
by the cosine of the mean angle $ \alpha (r) $, in the central region ($ r < 0.3\Rvir $).
In this way we fix the orientation, and assume the same rotational plane.
\begin{figure}
	\centering
	\subfloat[gas]{\includegraphics[width=0.50\textwidth] %
		{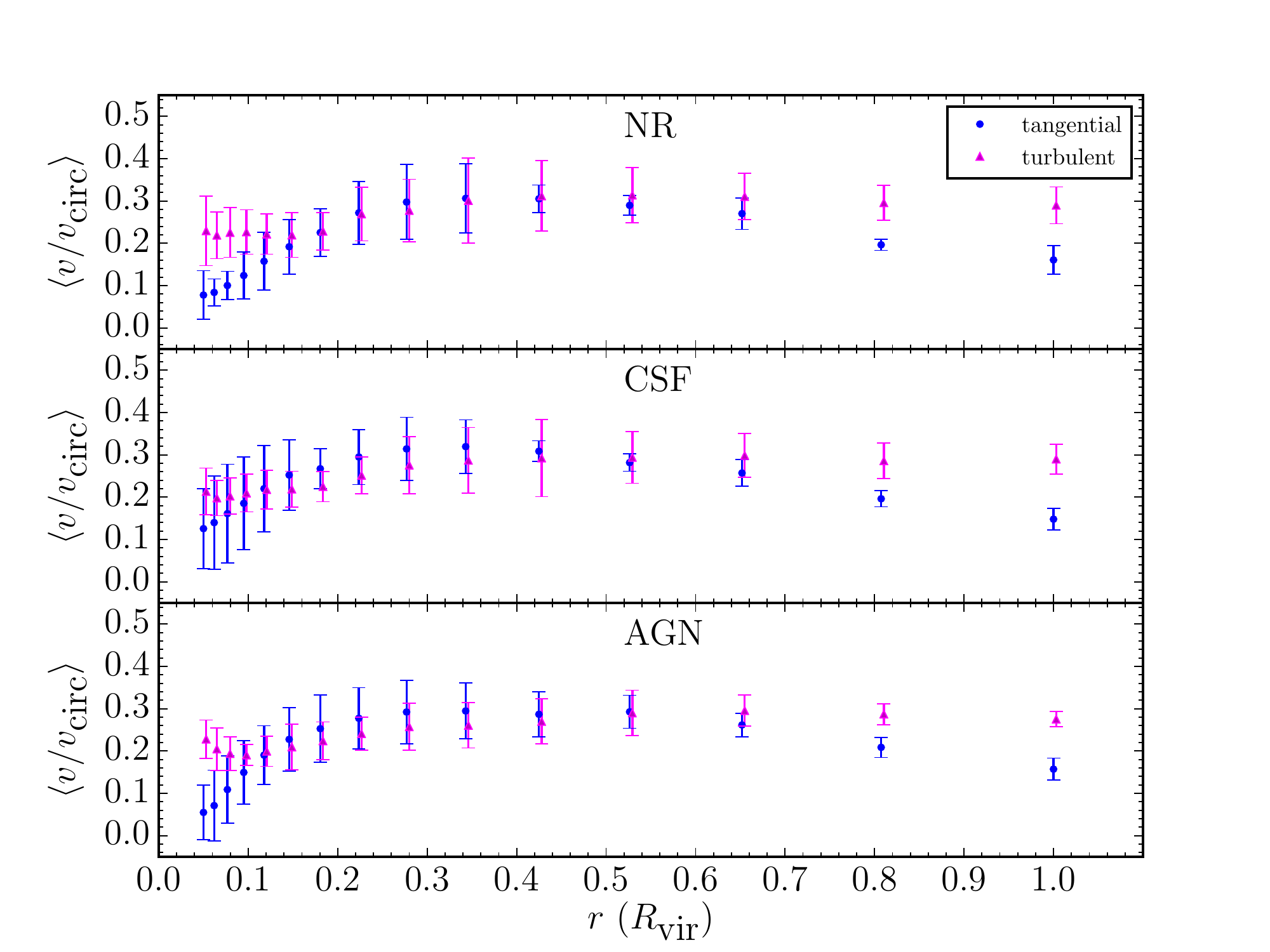}\label{fig:vtvturbgas}}
	\quad
	\subfloat[DM]{\includegraphics[width=0.50\textwidth] %
		{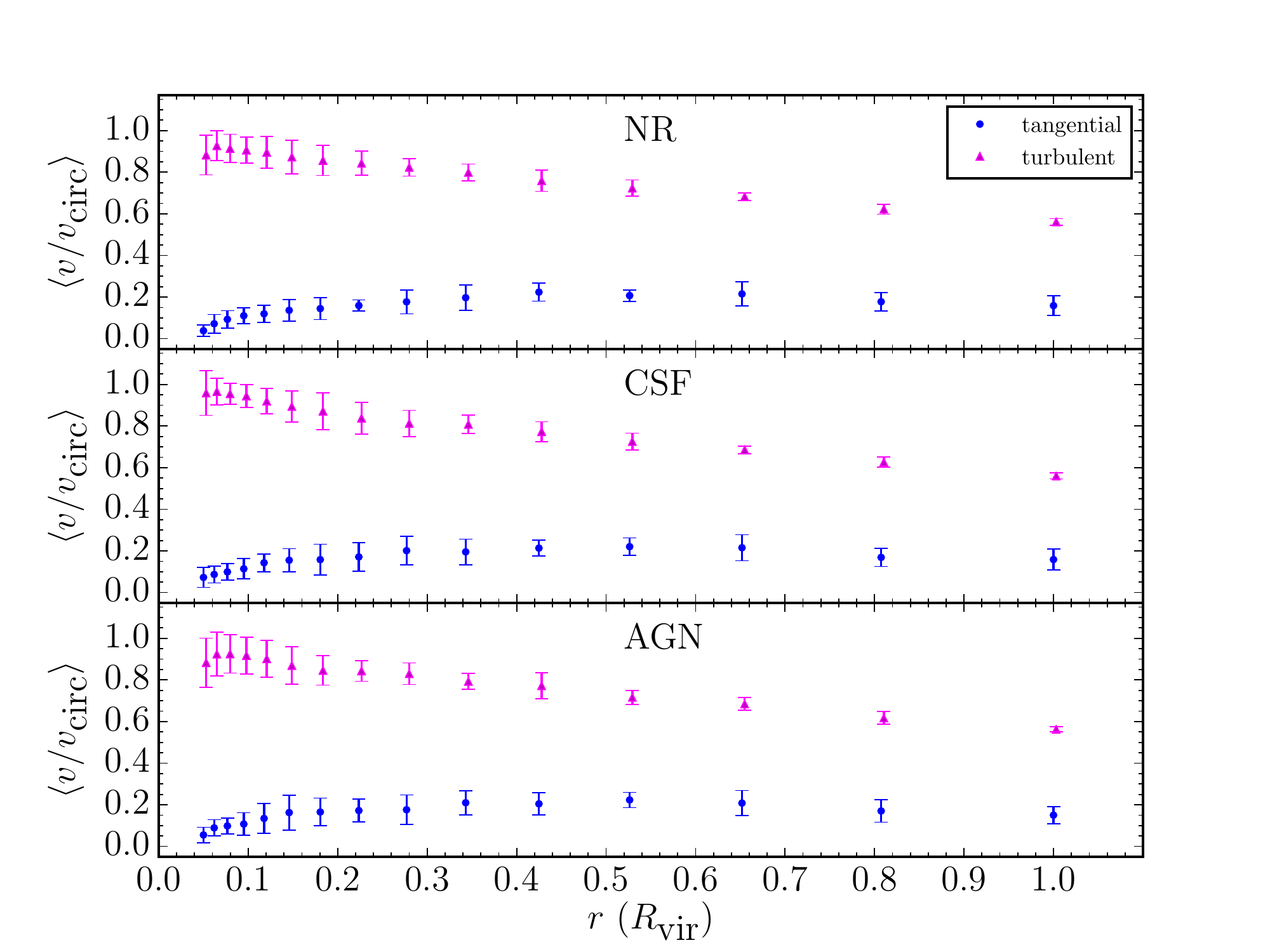}\label{fig:vtvturbDM}}
	\caption{\small Radial profiles of the mean tangential and turbulent velocity of ICM (upper panel)
		and DM (lower panel) for the rotating clusters only.
		The error bars indicate the standard deviation of the single profiles with respect to the mean profiles.}
\end{figure}
\begin{figure}
	\centering
	\subfloat[gas]{\includegraphics[width=0.50\textwidth] %
		{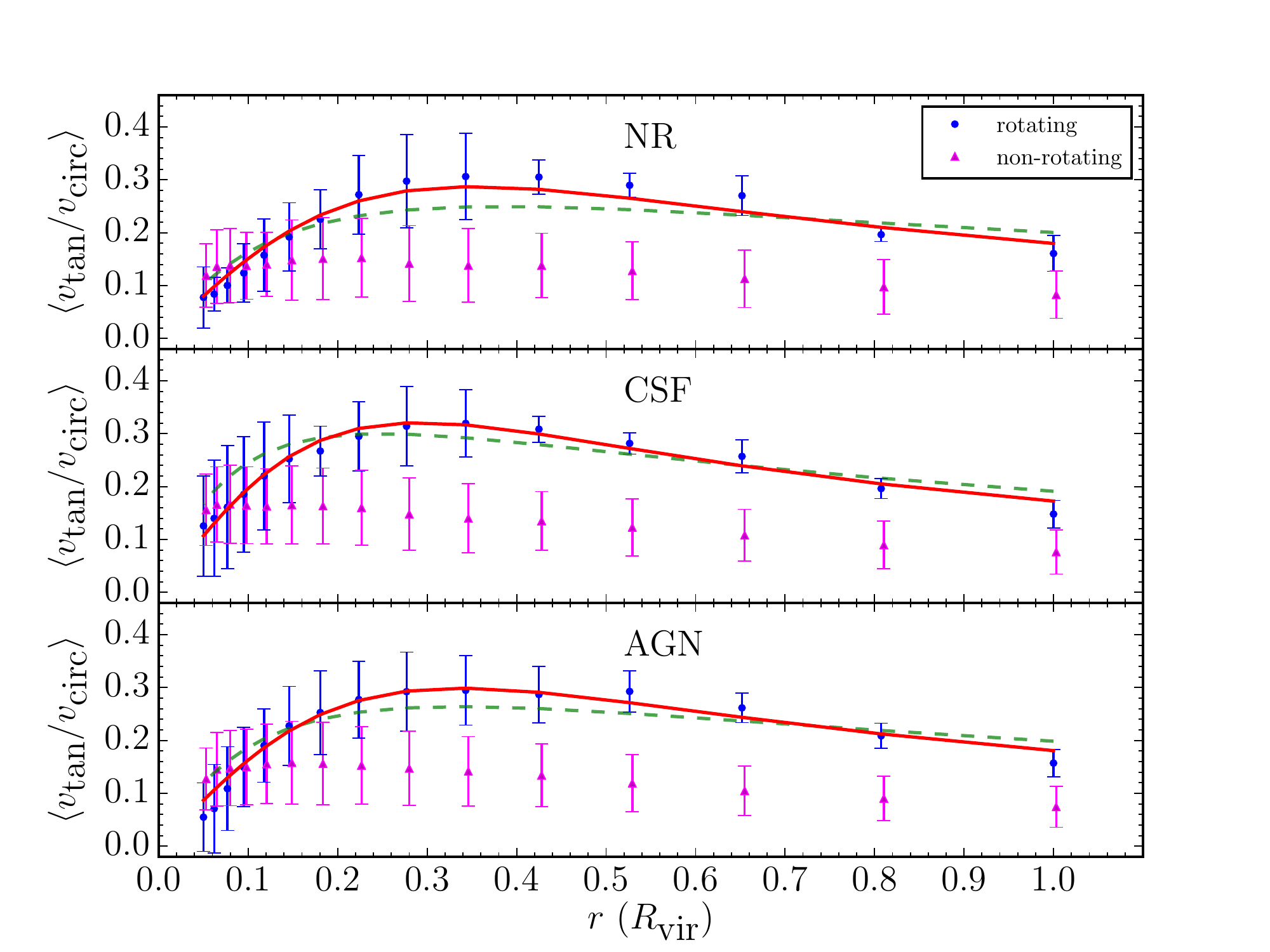}\label{fig:velocitiesgas}}
	\quad
	\subfloat[DM]{\includegraphics[width=0.50\textwidth] %
		{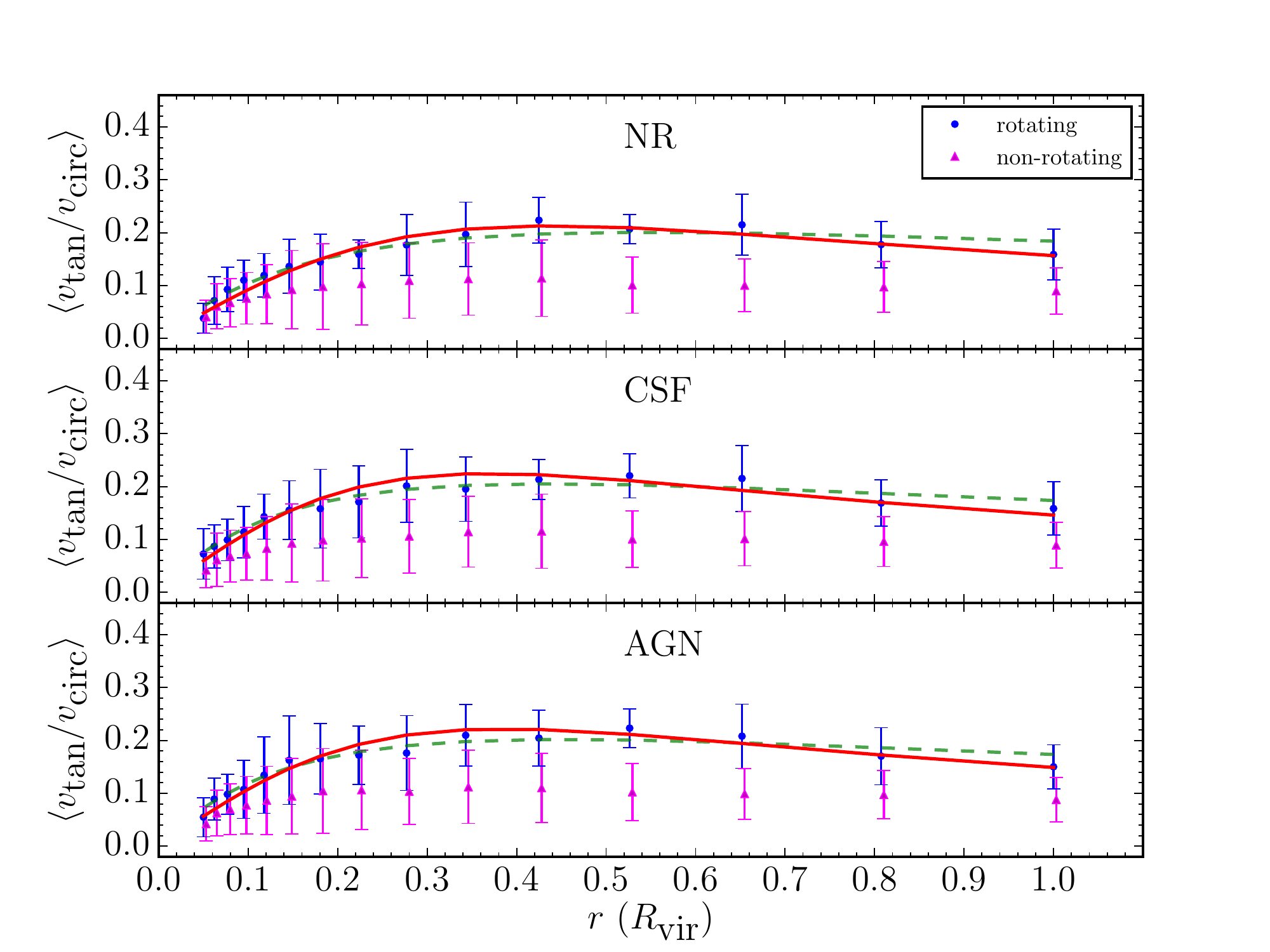}\label{fig:velocitiesDM}}
	\caption{\small Radial profiles of the mean rotational velocity of ICM (upper panel) and DM (lower panel)
		for the rotating and the non-rotating clusters.
		The error bars indicate the standard deviation of the single profiles with respect to the mean profiles.
		The dashed green line is the fit with the vp2 model, while the
		red solid line is the fit to the model of equation~\eqref{eqn:myvp2} (see text).}
	\label{fig:vt}
\end{figure}

By comparing the radial profiles of the mean tangential and turbulent velocity for the rotating clusters,
as shown in Fig.~\ref{fig:vtvturbgas} for the ICM and in Fig.~\ref{fig:vtvturbDM} for the DM, we find a
significant difference between these two matter components.
For the DM there is a net distinction between the two velocities: the turbulent velocity dominates
over the tangential one, with a decrease of $ \sim 30 $ per cent from the centre to the outskirts.
In the profiles of ICM velocities instead, there is a comparable contribution from turbulence
and coherent rotation in the region between $ \sim0.15 $ and $ \sim0.65\Rvir $.
Turbulence is still dominant in the innermost and in the outer regions, with a tendency to increase
for radial values between 0.1 and $ 0.4\Rvir $, and a flattening for higher radii.
Along the whole radial range, values vary between 0.2 and $ 0.3 v_\tup{circ} $,
corresponding to $ \sim 273 $ and $ \sim \SI{410}{\km\per\second} $.
The larger values of the DM velocity dispersion with respect to the gas can be explained in terms of
the absence of radiative mechanisms that remove kinetic energy of particles transforming it into thermal
energy, as in the case of gas particles.

From the comparison of our profiles of the velocity dispersion with other works, we find a general consistency.
In particular there is a fairly good agreement for the DM profiles,
that typically show a decreasing trend and have larger values with respect to the gas
\citep[see][]{sunyaev:turbulence,rasia:ICMDMdensity,faltenbacher:motions}.
A remarkable agreement can be found with \citet{ascasibar:vtan}, as the values
are compatible within the errors over all the considered radial range.
Values are generally around a thousand of \si{\km\per\second}
in the central regions, and differ more significantly in the outskirts.

The profiles of ICM turbulence show less regular behaviours.
A recurring trend is the flattening for radii $ r \gtrsim 0.75\Rvir $, and values typically span
a relatively narrow range. In particular our values are compatible with \citet{faltenbacher:motions}
in the innermost regions ($ r \sim 0.10\Rvir $), and with \citet{rasia:ICMDMdensity} and \citet{lau:motions}
at intermediate radii. The agreement with the latter is of particular interest, as they take into account the
dynamical state of the clusters, thus only the profiles of the relaxed ones have been compared here.
We find more marked differences with \citet{sunyaev:turbulence}, possibly because
only a cluster is considered in their analysis, thus they are more sensitive to single-cluster properties.

The radial profiles of the tangential velocity for the rotating and the non-rotating clusters are shown
in Fig.~\ref{fig:velocitiesgas} and in Fig.~\ref{fig:velocitiesDM}, for the ICM and the DM respectively.
Differently from the case of the turbulent velocity profiles, there is a common trend for both the gas and
the DM, consisting in the increase of the values in the innermost regions up to $ 0.3-0.4\Rvir $,
where they reach $ \sim \SI{400}{\km\per\second} $ for the ICM and $ \sim \SI{250}{\km\per\second} $ for
the DM, and a smooth decrease in the outskirts.
Values at virial radius are around 16 per cent of the circular velocity in the rotating clusters,
and 8 per cent in the non-rotating clusters (with no substantial differences between ICM and DM).
These results are in fairly good agreement with the values reported in \citet{ascasibar:vtan} for the DM
and in \citet{lau:motions} for the gas.
The plots in Fig.~\ref{fig:vt} clearly show that single profiles are affected by relatively large scatters,
because of the different intrinsic behaviours of individual clusters.

\begin{figure}
	\centering
	\includegraphics[width=0.50\textwidth]{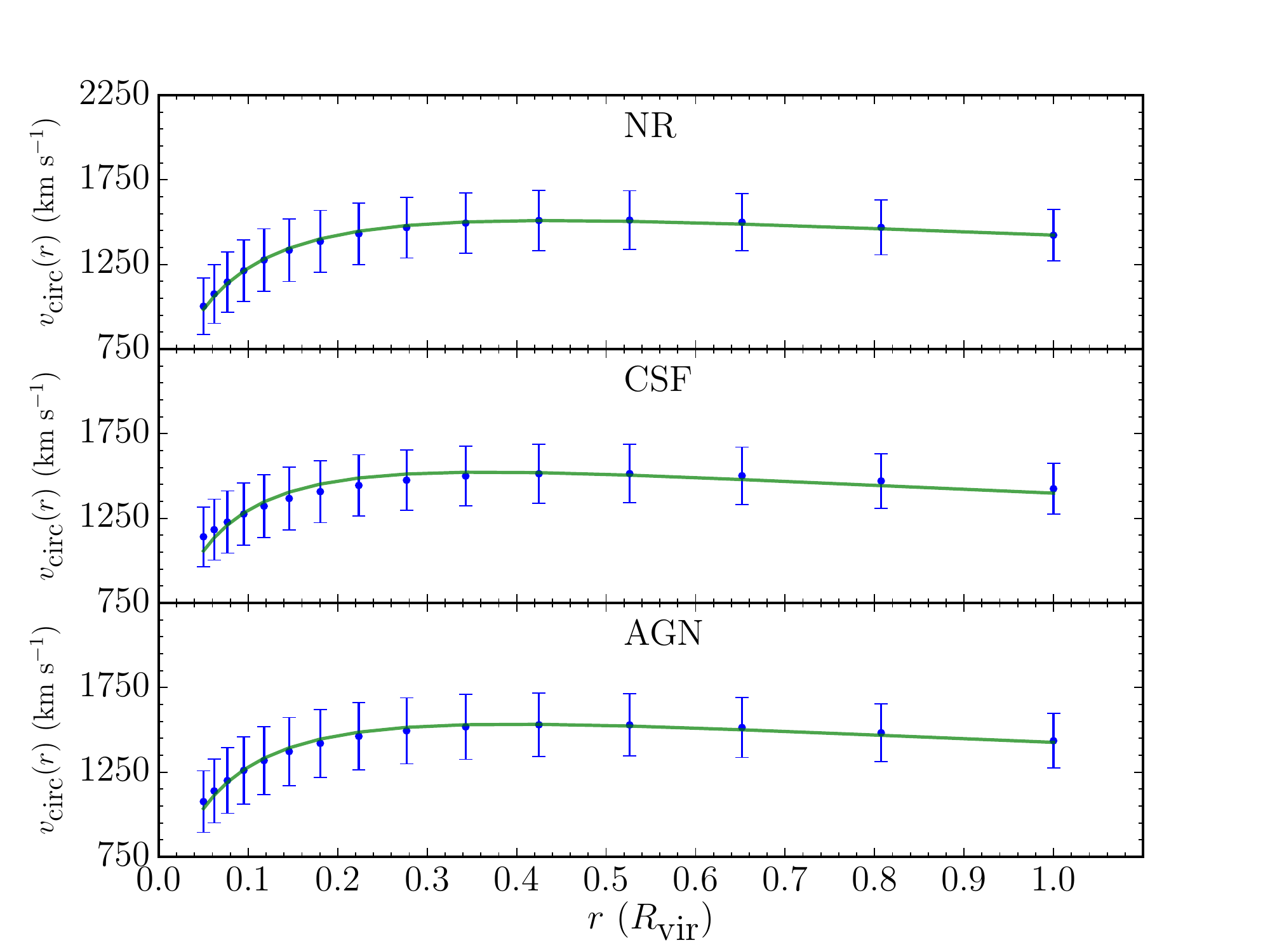}
	\caption{\small Circular velocity estimated inside spheres with increasing radius, averaged over all the clusters
		with $ 1 \sigma $ dispersion. The green solid line is the fit to the vp1 model of equation~\eqref{eqn:vp1}.}
	\label{fig:vcircr}
\end{figure}

In order to check whether our mean profiles of the tangential velocity can be described by an analytical
rotational model, we fit them to the two models introduced by \citet{bianconi:articolo}.
We neglected the simple solid body model, since we can see from the angular momentum profiles
shown in section~\ref{sec:angmomprofiles}
that it is not appropriate to describe the rotational motions in our cluster sample.
The two models that we consider refer to the case of a non-rigid rotating ICM, whose contribution to
the gravitational potential of the cluster is negligible.
The first proposed model, referred hereafter as vp1, is the circular velocity of the gas in a
Navarro-Frenk-White (NFW) DM density distribution \citep{nfw:article}, as a function of the radial
distance from the centre:
\begin{equation}
	\label{eqn:vp1}
	v_\tup{circ}(r) = v_{c0} \ \left[\frac{\ln(1 + r/r_0)}{r/r_0} - \frac{1}{1 + r/r_0}\right]^{\frac{1}{2}} \ ,
\end{equation}
where the radius $ r_0 $ corresponds to the peak value of the velocity.
Since it represents the circular velocity along the cluster radius, this profile is not fully appropriate to fit
our tangential velocity. Therefore we fit to vp1 the profile of the circular velocity computed over different radial spheres,
$ v_\tup{circ}(r) = \sqrt{G M(r)/r}$, and we find a very good agreement, as shown in Fig.~\ref{fig:vcircr}.
The second model proposed by \citet{bianconi:articolo} is an alternative to the circular velocity profile,
characterized by a steeper increase in the core regions and a deeper decrease in the outskirts:
\begin{equation}
	\label{eqn:vp2}
	v_\tup{tan}(r) = v_{t0} \ \frac{r/r_0}{(1 + r/r_0)^2}
\end{equation}
that will be referred hereafter as vp2 model.
We also introduce a modified version of vp2, the vp2b model, of equation:
\begin{equation}
\label{eqn:myvp2}
	v_\tup{tan}(r) = v_{t0} \ \frac{r/r_0}{1 + (r/r_0)^2} \ .
\end{equation}
The fits of the mean tangential velocity profiles to the vp2 and vp2b models
can be seen in Fig.~\ref{fig:velocitiesgas} for the gas and in Fig.~\ref{fig:velocitiesDM} for the DM.
The $ v_{t0} $ and $ r_0 $ parameters which best fit equations~\eqref{eqn:vp2} and~\eqref{eqn:myvp2}
are listed in Table~\ref{tab:velocityparameters}.
\begin{table}
	\centering
	\caption{\small Parameters of the fit with the vp2 (equation~\eqref{eqn:vp2}) and
		vp2b (equation~\eqref{eqn:myvp2}) models of the mean tangential velocity of gas and
		DM for rotating clusters.}
	\begin{tabular}{ccc}
		\toprule
		Dataset & $ (v_{t0} \pm \sigma_{v_{t0}}) \ (v_\tup{circ}) $ & \
			$ (r_0 \pm \sigma_{r_0}) \ (\Rvir) $\\
		\midrule
		\multicolumn{3}{c}{gas vp2}\\
		NR & $ 1.00 \pm 0.04 $ & $ 0.38 \pm 0.05 $ \\
		CSF & $ 1.21 \pm 0.07 $ & $ 0.24 \pm 0.04 $ \\
		AGN & $ 1.07 \pm 0.07 $ & $ 0.33 \pm 0.05 $ \\
		\multicolumn{3}{c}{DM vp2}\\
		NR & $ 0.80 \pm 0.06 $ & $ 0.55 \pm 0.12 $ \\
		CSF & $ 0.82 \pm 0.07 $ & $ 0.44 \pm 0.09 $ \\
		AGN & $ 0.81 \pm 0.07 $ & $ 0.45 \pm 0.10 $ \\
		\midrule
		\multicolumn{3}{c}{gas vp2b}\\
		NR & $ 0.58 \pm 0.03 $ & $ 0.35 \pm 0.03$ \\
		CSF & $ 0.65 \pm 0.04 $ & $ 0.29 \pm 0.03 $ \\
		AGN & $ 0.60 \pm 0.04 $ & $ 0.33 \pm 0.04 $ \\
		\multicolumn{3}{c}{DM vp2b}\\
		NR & $ 0.42 \pm 0.03 $ & $ 0.44 \pm 0.06$ \\
		CSF & $ 0.45 \pm 0.03 $ & $ 0.37 \pm 0.05 $ \\
		AGN & $ 0.44 \pm 0.04 $ & $ 0.38 \pm 0.06 $ \\
		\bottomrule
	\end{tabular}
	\label{tab:velocityparameters}
\end{table}
Both models are in agreement with the data within one standard deviation;
the residuals are lower for the vp2b, that better fits the ICM data, especially around the bump
observed at $ r \sim 0.3\Rvir $ and in the external regions.
The similar behaviours of gas and DM suggest a co-rotation of these
two components, that is further investigated in section~\ref{sec:componentsbehaviour}.

\section{Co-rotation of the DM and the ICM}
\label{sec:componentsbehaviour}
We drop here the distinction between relaxed and unrelaxed clusters, and we focus
on the specific angular momentum vectors of gas ($\bm j_\tup{gas}$)
and DM ($\bm j_\tup{DM}$) at virial radius, aiming at exploring possible correlations
between the two components.
In particular we compare the orientation and the absolute value of these two vectors.
Using the direction vectors, $ \hat j_\tup{gas} = \bm j_\tup{gas}/j_\tup{gas} $
and $ \hat j_\tup{DM}= \bm j_\tup{DM}/j_\tup{DM} $ (being $ j_\tup{gas} = \abs{\bm j_\tup{gas}} $
and $ j_\tup{DM} = \abs{\bm j_\tup{DM}} $),
the angle between the two angular momenta at virial radius is computed as
\begin{equation}
	\label{eqn:thetagasDM}
	\theta_\tup{gas,DM} = \arccos \left[\hat j_\tup{gas}(\Rvir) \cdot \hat j_\tup{DM}(\Rvir) \right] \ .
\end{equation}
For our goal, we consider that two vectors are aligned if $ \theta_\tup{gas,DM} < \SI{10}{\degree}$.
Under this condition the gas and DM particles are co-rotating, and the motions of DM could be inferred
by measuring the gas.

\begin{figure}
	\centering
	\includegraphics[width=0.50\textwidth] %
		{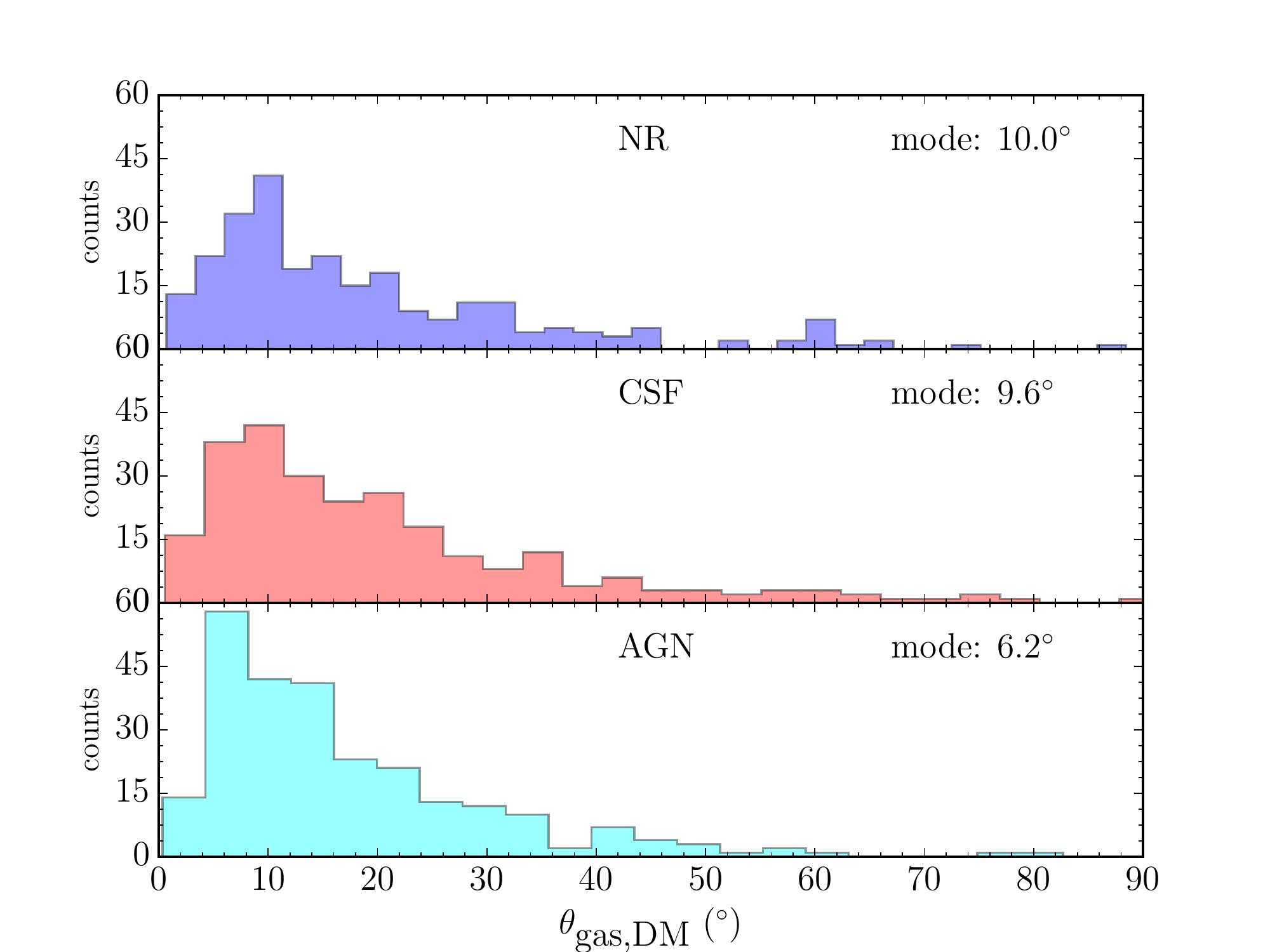}	
	\caption{\small Distributions of the angles between gas and DM angular momenta at
		virial radius. For $ \theta_\tup{gas,DM} > \SI{90}{\degree}$ there are only few isolated clusters.}
	\label{fig:thetagasDMhistograms}
\end{figure}

\begin{figure}
	\centering
	\includegraphics[width=0.50\textwidth] %
		{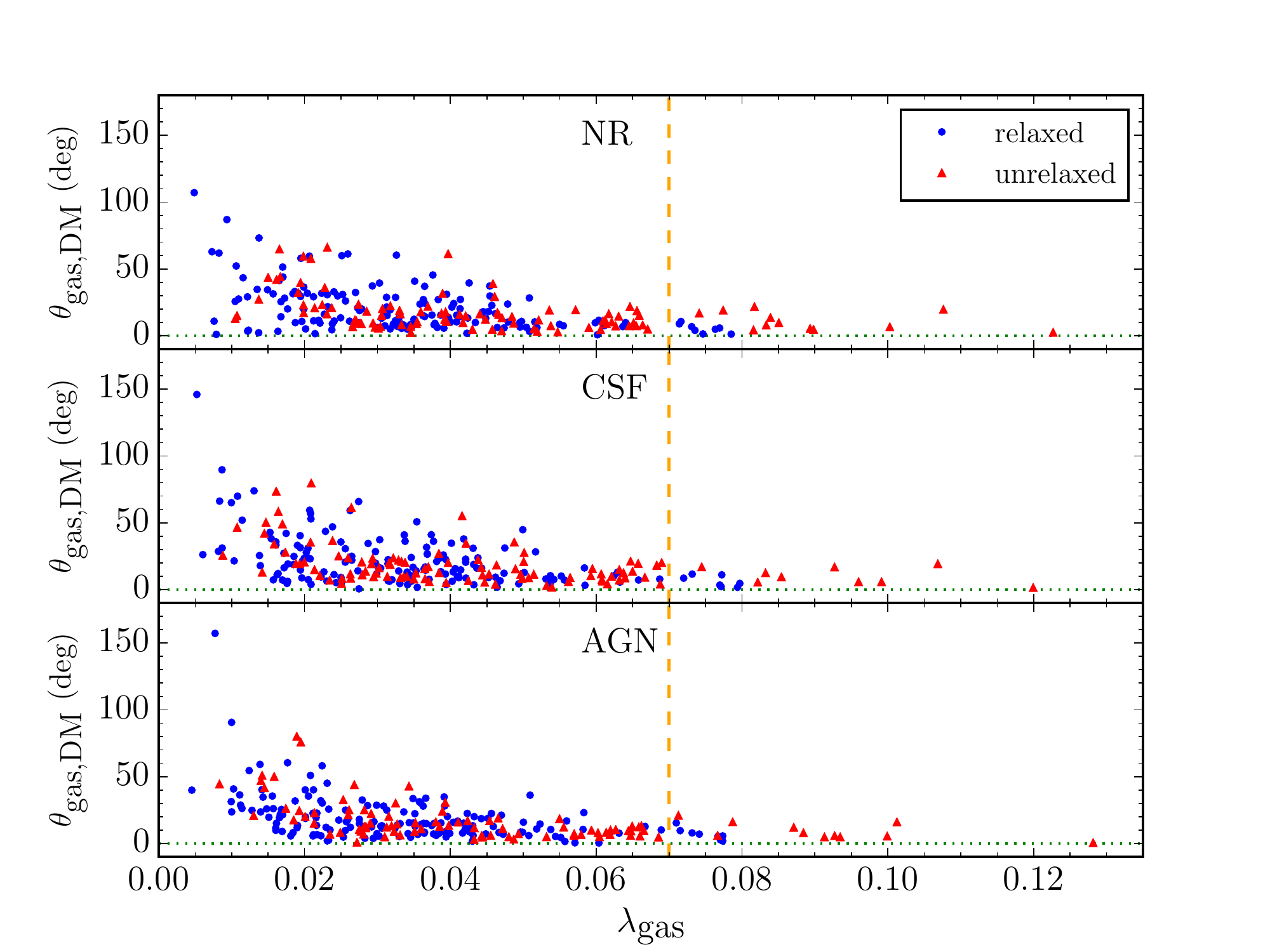}
	\caption{\small Angle $ \theta_\tup{gas,DM} $ versus the spin parameter of the gas,
		with the distinction for relaxed and unrelaxed clusters.
		The dashed orange line indicates the threshold value for the separation of the rotating clusters from the
		non-rotating ones.}
	\label{fig:thetalambdagas}
\end{figure}

The distribution of $ \theta_\tup{gas,DM} $ for all the clusters in the sample is reported in
Fig.~\ref{fig:thetagasDMhistograms}.
Around 40 per cent of the sample (corresponding to $ \sim $ 100 objects)
shows $ \theta_\tup{gas,DM} < \SI{10}{\degree}$.
In Fig.~\ref{fig:thetalambdagas} we plot the angle $ \theta_\tup{gas,DM} $ as a function of
$ \lambda_\tup{gas} $, discriminating by the relaxation state of the clusters.
We find that the dynamical state does not seem to play a relevant role on the alignment
between gas and DM.
The values of $ \theta_\tup{gas,DM} $ are below \SI{20}{\degree} for relatively high values
of $ \lambda_\tup{gas} $. In the clusters classified as rotating (having $ \lambda_\tup{gas} > 0.07 $)
the angle values are about \SI{10}{\degree}.
This leads to the conclusion that a larger cluster rotation is linked to a larger alignment of the angular momenta
of gas and DM. Such alignment can be seen as the evidence for a co-rotation of these
two components, considering that the orientation of the angular momentum for radial values
$ r \gtrsim 0.3\Rvir $ (see Fig.~\ref{fig:2rotgas_alpha}) is almost fixed.

\begin{figure}
	\centering
	\includegraphics[width=0.50\textwidth]{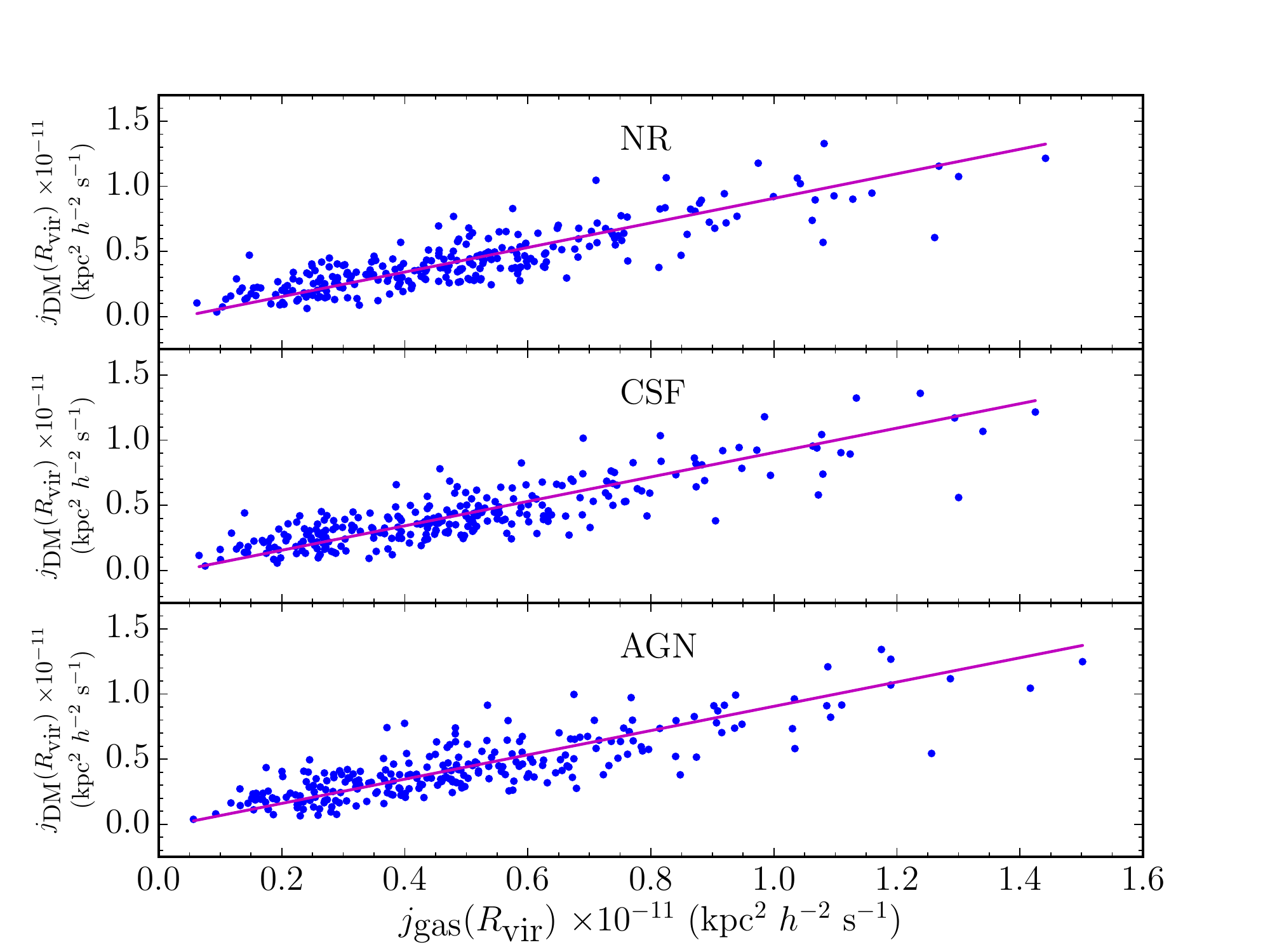}
	\caption{\small Relation between the absolute values of the specific angular momentum of DM and gas
		at virial radius; robust linear fits are also shown.}
	\label{fig:jDMjgas}
\end{figure}

As in the case of the orientation, we also expect a correlation in the absolute values of the angular momentum 
of the two components. Fig.~\ref{fig:jDMjgas} shows the relation between
$ j_\tup{gas}(\Rvir) $ and $ j_\tup{DM}(\Rvir) $ for our dataset.
A correlation between the absolute values is present, and the parameters of the linear fits to the data performed
with the bisector method are listed in Table~\ref{tab:jDMgasparameters}.
It is worth noting that the slope value of $ \sim 0.94 $ is consistent within the error
with the value obtained from the correlation between the spin parameters of the DM and gas component
(see Table~\ref{tab:lambdacompfit}).
From this result we find that the ICM specific angular momentum is a factor of $ \sim 1.06 $ larger than that of DM.
However, we also find that the gas angular momentum fraction
$ \ell_\tup{gas} = L_\tup{gas}/L_\tup{DM} \sim 0.17 $ at virial radius, meaning that when
masses are taken into account, the DM contribution to the angular momentum is dominant.
\begin{table}
	\centering
	\caption{\small Parameters of the robust linear fits performed on the $ j_\tup{DM}(\Rvir) $ vs
		$ j_\tup{gas}(\Rvir) $ plots. $ a $ and $ b $ are the slope and the zero-point respectively;
		$ c_\tup{corr} $ indicates the correlation coefficient.}
	\begin{tabular}{lccc}
		\toprule
		Dataset & $ c_\tup{corr}$ & $ a \pm \sigma_a $ & \
			$ b \pm \sigma_b \ \ (\si{\kpc\squared} h\minuss \si{\per\second})$\\
		\midrule
		NR & 0.85 & $ 0.94 \pm 0.04 $ & $(-4 \pm 2) \x \num{e-13} $\\
		CSF & 0.84 & $ 0.94 \pm 0.04 $ & $(-3 \pm 2) \x \num{e-13} $\\
		AGN & 0.83 & $ 0.93 \pm 0.04 $ & $(-3 \pm 2) \x \num{e-13} $\\
		\bottomrule
	\end{tabular}
	\label{tab:jDMgasparameters}
\end{table}
\section{Summary and conclusions}
\label{sec:conclusions}
In this work we have investigated the presence of rotational motions in the most massive clusters
of MUSIC-2, a large dataset of gas-dynamical simulations.
We considered different physical models to describe
the ICM: radiative and non-radiative, with and without AGN feedback. We focused on the study of the specific
angular momentum and of the tangential and random velocity components.
We have selected only the relaxed clusters, avoiding the impact of merging processes and/or large
substructures motions, which enlarge the contribution of turbulence.
We discriminate the rotating haloes through the value of the spin parameter of the gas.
It must be remarked that the threshold value that has been adopted is relatively high, in order to take into
account objects with a large rotational support.
This contributes in having a relatively low statistics of clusters classified as rotating.

The main results of this analysis can be summarized as follows.
\begin{itemize}
	\item In general, we find that the results are independent on the models 
	to describe the ICM properties (NR, CSF and AGN).
	This can be explained by the fact that the dynamics of clusters are dominated by the DM component;
	\item the spin parameter distributions for gas and DM in our sample of massive simulated clusters
	are consistent with previous results in literature;
	\item with the adopted criterion to discriminate the rotational state,
	we have 4 per cent of the relaxed clusters classified as rotating.
	The profiles of the specific angular momentum vector for ICM and DM along different radial distances
	from the centre, averaged over the aforementioned classes, show an almost constant orientation for the
	rotating clusters at radial distances larger than $ 0.3\Rvir $. The modulus of the specific angular momentum
	vector increases following a power-law behaviour with exponent $ \beta \sim 0.7 $ for gas and $ \beta \sim 1 $
	for DM, with the tendency to flatten in the outskirts. All these results lead to the conclusion
	that a solid body rotation model would not be correct for our clusters;
	\item the average profiles of the tangential velocity show that the rotational support is small relative to
	the circular velocity at virial radius ($ \sim $ 16 per cent for the rotating clusters), but yet non-zero.
	The velocity dispersion generally dominates, especially for DM.
	The behaviour of the tangential velocity profile of the rotating clusters, can be modelled with a simple
	modification of the circular velocity profile derived from the NFW dark matter density distribution in the haloes;
	\item in a non-negligible fraction of clusters ($ \sim 40 $ per cent of the total sample) there is evidence
	for a common behaviour of the DM and the gas components, since their specific angular momenta are
	correlated both in direction and in modulus. This indicates a possible co-rotation, that is suggested
	also by the comparison of the behaviour of the radial profiles.
\end{itemize}

The proof of DM and ICM co-rotation is one of the key results of this study as it can lead to the possibility of
inferring DM motions by studying gas motions.
They are definitely challenging to measure, however
a variety of observational techniques at different wavelengths can be used.
Among these, X-ray spectroscopy (or surface brightness mapping) and SZ mapping are the most promising.
In particular, instruments with high angular resolutions are necessary for both the production of
maps of the kinematic SZ effect towards clusters at a frequency around \SI{200}{\GHz}, and spectroscopic
measurements on emission lines from heavy elements in the ICM. The NIKA2 camera at the IRAM 30-m telescope
\citep{monfardini:nika1e2} and the \textit{Athena} satellite \citep{athena:presentation} respectively, could satisfy
these requirements.

In a companion paper we produce synthetic maps of kinematic SZ effect, with which it is possible to map
the velocities along the line of sight, and to determine the presence of a rotational motion, if any.
We also plan to investigate the correlation between the dynamical properties of ICM and galaxies in our sample,
in order to compare the results with the study of cluster rotation inferred from the velocity of galaxies.
In particular we aim to apply the analysis developed in \citet{plionis:rotation} to our sample of clusters to study
possible correlations.

\section*{Acknowledgements}
The authors thank the anonymous referee for the constructive comments on the text.
They also want to thank Veronica Biffi, Stefano Borgani, Giuseppe Murante and Susanna Planelles
for their contribution to the implementation of the code for the run with the AGN feedback and for their
comments and suggestions.
The MUSIC simulations have been performed in the Marenostrum supercomputer at the
Barcelona Supercomputing Centre, thanks to the computing time awarded  by
Red Espa\~{n}ola de Supercomputaci\'on.
This work has been partially supported by funding from Sapienza University of Rome - Progetti di Ricerca Anno 2014
prot. C26A14KYYJ and Anno 2015 prot. C26A15LXNR.
GY and FS acknowledge financial support from MINECO (Spain) under research grants  AYA2012-31101 and
AYA2015-63810-P.
ER is financially supported by PIIF-GA-2013-627474 and NSF AST-1210973.

\bibliographystyle{mnras}
\bibliography{./BIBLIOGRAPHY.bib}

\appendix
\section{Critical value of the gas spin parameter for rotating clusters}
\label{sec:appendix}
Since there is not a universal critical value for $ \lambda_\tup{gas} $ that can be adopted to discriminate
rotating objects, we choose the threshold by inspecting the radial average profiles of the tangential velocity
(see the detailed description in section~\ref{sec:velocities}) of the two populations of rotating and
non-rotating clusters, ($ v_\tup{tan}^{rot}(r) $ and $ v_\tup{tan}^{nonrot}(r) $, respectively).
We take the value for which these profiles are separated
more than the corresponding standard deviations
over $ r \gtrsim 0.3\Rvir $, that is the radial range where the angular momentum orientation is almost
fixed (see section~\ref{sec:angmomprofiles}).
To quantify the separation of the profiles at a radius $ r $, indicating with $ \sigma_\tup{tan}^{rot}(r) $ and
$ \sigma_\tup{tan}^{nonrot}(r) $ the corresponding standard deviations
(represented by the error bars in the profile plots), we introduce the following estimator
\begin{equation}
	d_v(r) = \frac{\abs{v_\tup{tan}^{rot}(r) - v_\tup{tan}^{nonrot}(r)}} %
		{\sigma_\tup{tan}^{rot}(r)+\sigma_\tup{tan}^{nonrot}(r)} \ ,
\end{equation}
so that they can be considered as separated when $ d_v(r) > 1 $.
The best $ \lambda_\tup{gas,crit} $ is the one for which the minimum value of $ d_v(r) $,
$ d_{vm} $, is larger than one in the range $ r \gtrsim 0.3 \Rvir $.
The fraction of relaxed clusters which, according to our criterion, can be defined as rotating is listed in
Table~\ref{tab:lambdadata} for some values of $ \lambda_\tup{gas,crit} $, together with
$ d_{vm}$.
\begin{table}
	\centering
	\caption{\small Threshold values for the gas spin parameter and corresponding percentage of rotating clusters
		with respect to the number of relaxed clusters, $ N_\tup{rot}/N_\tup{rel} $.
		The $ d_{vm} $ value is also shown (see text).}
	\begin{tabular}{ccc}
		\toprule
		$  \lambda_\tup{gas,crit} $ & $ N_\tup{rot}/N_\tup{rel} $ & $ d_{vm} $ \\
		\midrule
		0.03 & 49\%& 0.49 \\
		0.05 & 10\% & 0.85 \\
		0.07 & 4\% & 1.11 \\
		\bottomrule
	\end{tabular}
	\label{tab:lambdadata}
\end{table}
It results that the $ \lambda_\tup{gas,crit} $ having $ d_{vm} > 1 $ in the chosen radial range is 0.07,
therefore we adopt this value as the discriminating one.
\begin{figure}
	\centering
	\includegraphics[width=0.50\textwidth]{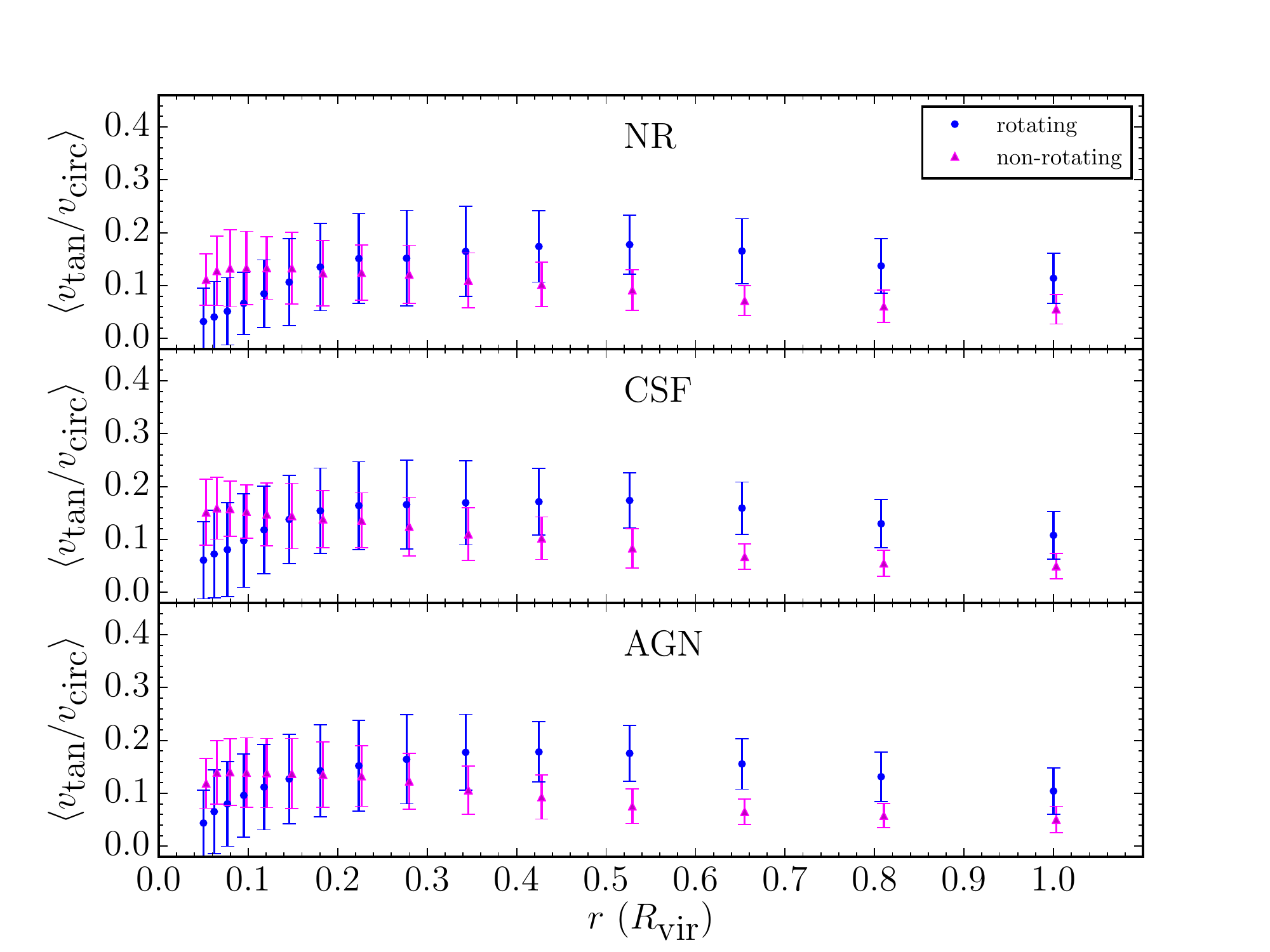}
	\caption{\small Radial profiles of the tangential velocity of the gas for the rotating and the non-rotating
		clusters, assuming $ \lambda_\tup{gas,crit} = 0.03 $.
		See the caption of Fig.~\ref{fig:vt} for a detailed description of the plots.}
	\label{fig:otherlambdas}
\end{figure}
Fig.~\ref{fig:otherlambdas} shows the profiles for $ \lambda_\tup{gas,crit} = 0.03 $,
where the overlapping of the two classes for $ r \lesssim 0.5 \Rvir $ is evident.
Values of $ \lambda_\tup{gas,crit} $ larger than 0.07 cannot be tested,
since the maximum spin parameter of the gas in the sub-sample of
relaxed clusters is $ \sim 0.078 $. From Fig.~\ref{fig:lambdadistributions} it can be seen that these values
correspond to the tails of the spin parameter distributions.

\bsp	
\label{lastpage}
\end{document}